\documentclass[pre,aps,twocolumn,superscriptaddress]{revtex4-1}
\usepackage{graphicx}
\usepackage{amssymb,amsfonts,amsmath}
\usepackage{color}
\usepackage{ulem}
\usepackage[hidelinks]{hyperref}
\usepackage{bbm}
\usepackage{bbold}

\usepackage{tikz}
\usetikzlibrary{shapes}
\usetikzlibrary{patterns}
\usetikzlibrary{angles, quotes}

\newcommand{\rv}{{\mathbf r}}

\newcommand{\ev}{{\bf e}}

\newcommand{\Jv}{{\bf J}}

\newcommand{\Fv}{{\bf F}}
\newcommand{\fv}{{\bf f}}

\newcommand{\vel}{{\bf v}}

\newcommand{\ms}{}

\newcommand{\msphantom}[1]{$\ldots$}

\newcommand{\eps}{{\boldsymbol \epsilon}}

\newcommand{\ad}{{\rm ad}}

\newcommand{\ext}{{\rm ext}}
\newcommand{\exc}{{\rm exc}}
\newcommand{\rt}{(\rv,t)}

\renewcommand{\sup}{{\rm sup}}
\newcommand{\str}{{\rm str}}
\newcommand{\flow}{{\rm flow}}
\newcommand{\eqr}[1]{Eq.~\eqref{#1}}
\newcommand{\DL}{\star}

\begin{document}

\title{Perspective: How to overcome dynamical density functional theory}

\author{Daniel de las Heras}
\affiliation{Theoretische Physik II, Physikalisches Institut, 
  Universit{\"a}t Bayreuth, D-95447 Bayreuth, Germany}

\author{Toni Zimmermann}
\affiliation{Theoretische Physik II, Physikalisches Institut, 
  Universit{\"a}t Bayreuth, D-95447 Bayreuth, Germany}

\author{Florian Samm\"uller}
\affiliation{Theoretische Physik II, Physikalisches Institut, 
  Universit{\"a}t Bayreuth, D-95447 Bayreuth, Germany}

\author{Sophie Hermann}
\affiliation{Theoretische Physik II, Physikalisches Institut, 
  Universit{\"a}t Bayreuth, D-95447 Bayreuth, Germany}

\author{Matthias Schmidt}
\affiliation{Theoretische Physik II, Physikalisches Institut, 
  Universit{\"a}t Bayreuth, D-95447 Bayreuth, Germany}
\email{Matthias.Schmidt@uni-bayreuth.de}

\date{28 January 2023, \ms revised version: 23 March 2023}

\begin{abstract}
We argue in favour of developing a comprehensive dynamical theory for
rationalizing, predicting, {\ms designing}, and machine learning
nonequilibrium phenomena that occur in soft matter. To give guidance
for navigating the theoretical and practical challenges that lie
ahead, we discuss and exemplify the limitations of dynamical density
functional theory. Instead of the implied adiabatic sequence of
equilibrium states that this approach provides as a makeshift for the
true time evolution, we posit that the pending theoretical tasks lie
in developing a systematic understanding of the dynamical functional
relationships that govern the genuine nonequilibrium physics. While
static density functional theory gives a comprehensive account of the
equilibrium properties of many-body systems, we argue that power
functional theory is the only present contender to shed similar
insights into nonequilibrium dynamics, including the recognition and
implementation of exact sum rules that result from the Noether
theorem.  As~a~demonstration of the power functional point of view, we
consider an idealized steady sedimentation flow of the
three-dimensional Lennard-Jones fluid and machine-learn the kinematic
map from the mean motion to the internal force field.  {\ms The
  trained model is capable of both predicting and designing the steady
  state dynamics universally for various target density
  modulations. This demonstrates the significant potential of using
  such techniques in nonequilibrium many-body physics and overcomes
  both the conceptual constraints of dynamical density functional
  theory as well as the limited availability of its analytical
  functional approximations.}
\end{abstract}

\maketitle

\section{Introduction}
\label{SECintroduction}
The coupled dynamics of the microscopic degrees of freedom in typical
soft matter systems generates a wide array of relevant and also often
unsolved nonequilibrium phenomena
\cite{nagel2017,evans2019physicsToday}. One central quantity for the
characterization of self-assembly and structure formation in complex
systems is the microscopically resolved one-body density distribution
$\rho\rt$, where~$\rv$ indicates position and $t$ denotes time. The
``density profile'' $\rho\rt$ acts as a central order parameter both
due to its intuitive physical interpretation and clearcut mathematical
definition~\cite{hansen2013}.

According to the {\it dynamical density functional theory} (DDFT), as
originally proposed by Evans in 1979 \cite{evans1979} {\ms and put at
  center stage by Marconi and Tarazona in 1999 \cite{marconi1999}},
the time evolution of the microscopic density profile is assumed to be
determined by the following partial differential equation:
\begin{align}
  \frac{\partial \rho(\rv,t)}{\partial t}
  &= \gamma^{-1} \nabla  \cdot  \rho(\rv,t)
  \nabla  \Big(
  \frac{\delta F[\rho]}{\delta \rho(\rv,t)} + V_{\rm ext}(\rv,t)
  \Big).
  \label{EQddft}
\end{align}
Here $\gamma$ is a friction constant, $F[\rho]$ is an intrinsic free
energy functional that depends functionally on the density profile,
and the external potential $V_\ext(\rv,t)$ represents interactions of
the system with the environment. The system is set into motion by a
temporal variation of $V_\ext\rt$, such as e.g.\ step-like switching
at an initial time.

The time evolution according to \eqr{EQddft} conserves the particle
number locally and hence it constitutes dynamics of model B type
\cite{hohenberg1977}.  In standard applications one starts with an
equilibrium state of the system and then the dynamics are monitored on
the basis of numerical time integration of Eq.~\eqref{EQddft}, {\ms
  where the time dependence is induced by the temporal variation of
  $V_\ext\rt$.} In order to provide reference data and to allow for
the generation of benchmark results to assess the quality of the
theory, resorting to many-body computer simulations is common, with
overdamped Brownian dynamics (BD) being a popular choice.  {\ms
  Ref.~\cite{marconi1999} initially spelled out the connection of
  these dynamics with DDFT} and Ref.~\cite{sammueller2021} describes a
modern and stable adaptive time-stepping BD simulation
algorithm. Comparison of DDFT data with experimental results are more
scarce, but notable exceptions include non-equilibrium sedimentation
of colloids \cite{royall2007dynamicSedimentation}, the self-diffusion
of particles in complex fluids \cite{bier2008prl}, the bulk dynamics
of Brownian hard disks~\cite{stopper2018dtpl}, {\ms and the flow
  profile and drying pattern of dispersion droplets \cite{perez2021}.}

The DDFT time evolution reaches a stationary state if the gradient on
the right hand side of \eqr{EQddft} vanishes, i.e.\ provided that the
expression inside of the parentheses is constant:
\begin{align}
  \frac{\delta F[\rho]}{\delta \rho(\rv)} + V_{\rm ext}(\rv) = \mu.
  \label{EQdft}
\end{align}
Here we have dropped the dependence on time in the notation, as the
situation is now static. The constant $\mu$ can be identified with the
chemical potential, which in a grand canonical statistical mechanical
setting is the conjugate control parameter of the mean particle
number. Equation \eqref{EQdft} is exact in equilibrium, as was shown
by Evans \cite{evans1979}. He proved the equilibrium intrinsic free
energy functional $F[\rho]$ to exist, to be unique, and to form the
starting point for a modern equilibrium theory of spatially
inhomogeneous liquids and crystals \cite{evans1992,evans2016}.

In practice one needs to rely on approximations for~$F[\rho]$, given a
microscopic fluid model under consideration. Once one has solved
\eqr{EQdft} for given values of~$\mu$ and temperature $T$ (the
dependence of $F[\rho]$ on $T$ is suppressed in the notation), then in
principle complete knowledge of the thermal system is available. The
value of the density functional $F[\rho]$ is the true intrinsic free
energy, and higher-order correlation functions are determined via
higher-order derivatives of the free energy functional or via
test-particle procedures. In particular two-body correlations
functions, such as the bulk pair correlation function $g(r)$ as well
as its generalization to inhomogeneous systems are accessible. These
exhibit defining characteristics of liquids and more general soft
matter systems and they are formally fully contained in the static
density functional theory framework.

Together with a number of available reliable approximate free energy
functionals, density functional theory is a powerful theoretical
framework that has been used to elucidate much intricate and complex
behaviour in soft matter. Recent representative highlights include
tracing hydrophobicity to critical drying at substrates
\cite{evans2019pnas,coe2022prl,coe2023}, resolving three-dimensional
structures of electrolyte aqueous solutions near surfaces
\cite{martinjimenez2017natCom,hernandez-munoz2019}, and addressing the
magnitude of the decay lengths in electrolytes
\cite{cats2021decayLength}. Rosenfeld's celebrated hard sphere
fundamental measure free energy functional \cite{rosenfeld1989,
  roth2010,roth2002WhiteBear,roth2006WhiteBear} is at the core of much
of this research activity.

In the following we wish to address whether or not the DDFT has the
prowess to play a similar role in nonequilibrium, as is often at least
implicitly assumed. We demonstrate on the basis of an explicit and
generic example, i.e., that of uniaxial compressional flow of the
three-dimensional Lennard-Jones fluid, that the DDFT is fundamentally
flawed and that in reality, as represented by many-body simulations,
recognizing the flow field as a further relevant degree of freedom is
required to represent true nonequilibrium. These conclusions are based
on analytical power functional approximations, adaptive BD simulation
data, and explicit machine learning of the power functional map from
motion to the interparticle one-body force field. {\ms Neglecting the
  dependence on the velocity field, via artificially setting its value
  identically to zero, reduces to the machine-learned functional
  mapping and hence the adiabatic time evolution of DDFT, albeit here
  on the basis of the quasi-exact adiabatic forces as they are
  included in the supervised machine learning.}

This Perspective is organized as follows. We first make some key
aspects of DDFT explicit in Sec.~\ref{SEClimited} and describe several
prominent shortcomings of this theory. We then give an account of how
to go towards the formally exact one-body dynamics in
Sec.~\ref{SEConeBody} and provide in Sec.~\ref{SECpft} a description
of key aspects of the power functional framework, which as we wish to
argue overcomes the fundamental defects of DDFT. We describe the
exemplary stationary compressional flow situation in
Sec.~\ref{SECsteadyStates} and lay out the application of Noether's
theorem in this statistical mechanical setting in
Sec.~\ref{SECnoether}. We present machine learning results for the
kinematic functional relationships of the streaming Lennard-Jones
fluid in Sec.~\ref{SECmachineLearning}. {\ms This method also yields
  direct access to the adiabatic force field, as is required for the
  DDFT time evolution, without the need for involving any prior
  explicit analytical approximations for the free energy density
  functional.}  We give conclusions and an outlook in
Sec.~\ref{SECconclusions}. {\ms Readers who are primarily interested
  in the machine learning aspects of our work
  (Sec.~\ref{SECmachineLearning}) are welcome to skip to
  appendix~\ref{SECappendix} where we lay out our strategy of its use
  in predicting and designing nonequilibrium many-body dynamics in
  soft matter.}

\section{limits and limitations of adiabatic dynamics}
\label{SEClimited}
We go into some detail and describe why the DDFT represents adiabatic
dynamics in the sense of a temporal sequence of spatially
inhomogeneous equilibrium states. The equilibrium intrinsic free
energy functional splits into ideal and excess (over ideal gas)
contributions according to $F[\rho]=F_{\rm id}[\rho]+F_{\rm
  exc}[\rho]$. Here the excess free energy functional $F_{\rm
  exc}[\rho]$ accounts for the effects of the interparticle
interactions on the equilibrium properties of the system and it is in
general unknown and requires approximations to be made.  The ideal gas
free energy functional however is exactly given by
\begin{align}
  F_{\rm id}[\rho] &= k_BT
  \int d\rv \rho(\rv)[\ln(\rho(\rv)\Lambda^3)-1],
\end{align}
where $k_B$ denotes the Boltzmann constant, $\Lambda$ is the thermal
de Broglie wavelength, and we consider three-dimensional systems.  The
functional derivative, as it is relevant for \eqr{EQddft}, is $\delta
F_{\rm id}[\rho]/\delta \rho(\rv)=k_BT \ln(\rho(\rv)\Lambda^3)$. When
disregarding the excess contribution and inserting this result alone
into the DDFT equation of motion \eqref{EQddft}, its right hand side
becomes $\gamma^{-1} \nabla\cdot \rho(\rv,t) \nabla [k_BT
  \ln(\rho(\rv,t)\Lambda^3) + V_\ext\rt]$ {\ms with the dependence on
  $\Lambda$ being irrelevant due to the spatial gradient
  operation. The result} can be re-written further such that for the
case of the ideal gas, where $F_{\rm exc}[\rho]=0$ and $F[\rho]=F_{\rm
  id}[\rho]$, the equation of motion \eqref{EQddft} attains the
following form:
\begin{align}
  \frac{\partial\rho(\rv,t)}{\partial t} &= 
  D_0 \nabla^2 \rho\rt 
  - \nabla \cdot \rho\rt \fv_\ext\rt/\gamma.
  \label{EQidealDiffusionEquation}
\end{align}
Here $D_0=k_BT/\gamma$ is the diffusion constant, $\nabla^2$ is the
Laplace operator and the external force field is given (here) as
$\fv_\ext\rt=-\nabla V_\ext\rt$. Equation
\eqref{EQidealDiffusionEquation} is the exact drift-diffusion equation
for overdamped motion of a mutually noninteracting system, i.e., the
ideal gas.

Besides Evans' original proposal \cite{evans1979} based on the
continuity equation and undoubtedly his physical intuition,
derivations of the DDFT \eqref{EQddft} were founded much more recently
on Dean's equation of motion for the density operator
\cite{marconi1999}, the Smoluchowski equation \cite{archer2004}, a
stationary action principle for the density \cite{chan2004}, the
projection operator formalism \cite{espanol2009}, a phase-space
approach \cite{marconi2007}, the mean-field approximation
\cite{dzubiella2003mfddft}, a local equilibrium assumption
\cite{lutsko2021reconsidering}, and a non-equilibrium free energy
\cite{szamel2022}. The question of the well-posedness of the DDFT was
addressed \cite{goddard2021wellposedness} and several extensions
beyond overdamped Brownian dynamics were formulated, such as e.g.\ for
dynamics including inertia \cite{archer2006inertia, archer2009inertia,
  goddard2012prl, stierle2021} and for particles that experience
hydrodynamic interactions~\cite{rex2009epje,goddard2012prl} or undergo
chemical reactions \cite{monchojorda2020,bley2021}.

The DDFT was also used beyond the description of fluids, such as
e.g.\ for opinion dynamics~\cite{goddard2021opinion} and epidemic
spreading \cite{tevrugt2020natComm}. Recent reviews of DDFT are given
in Refs.~\cite{tevrugt2020review,tevrugt2022perspective}, {\ms with
  Ref.~\cite{tevrugt2022perspective} giving an updated overview of
  several very recent directions.}  The theory is put into a wider
perspective, together with much background pedagogical material in
Ref.~\cite{schmidt2022rmp}.  A modern and well-accessible account of
the general strategy of dynamical coarse-graining in statistical
physics, of which the DDFT can be viewed as being a representative,
has recently been given by Schilling~\cite{schilling2022}.

The fact that both the static limit for the fully interacting system
\eqref{EQdft} as well as the full dynamics of the noninteracting
system \eqref{EQidealDiffusionEquation} are exact, taken together with
the heft of the DDFT literature, appears to give much credibility to
the equation of motion \eqref{EQddft}. However, despite the range of
theoretical techniques employed \cite{marconi1999, archer2004,
  chan2004, espanol2009, marconi2007, dzubiella2003mfddft, szamel2022,
  lutsko2021reconsidering} neither of these approaches has provided us
with a concrete way of going beyond \eqr{EQddft}. Apart from several
case-by-case and rather {\it ad hoc} modifications, no systematic or
even only practical identification of what is missing has been
formulated.  (We turn to power functional theory in
Sec.~\ref{SECpft}.)  This is a problematic situation as two defects of
\eqr{EQddft} are immediately obvious upon inspection: i) the
description is local in time and there is no natural mechanism for the
inclusion of memory while time-locality is not sufficient for general
nonequilibrium situations; ii)~only flow that leads to direct changes
in the density profile is captured and hence effects of rotational
flow, such as shearing, as well as of nonequilibrium effects in
compression and expansion are lost (see below).

Here we argue that these defects are indicative of a broader failure
of \eqr{EQddft} to describe nonequilibrium physics. We show that the
DDFT is only fit to describe situations in which the dynamics follow
an adiabatic path through a sequence of equilibrium states. The
description of genuine nonequilibrium dynamics in a functional setting
on the one-body level rather requires recognition of the local
velocity field as a further relevant physical variable besides the
density profile, and this is provided by power functional theory
\cite{schmidt2022rmp}. Before laying out key principles of this
approach in Sec.~\ref{SECpft}, we first describe the microscopically
sharp coarse-graining on the one-body level of correlation functions.

\section{Towards exact one-body dynamics}
\label{SEConeBody}
Evans based his original derivation \cite{evans1979} of \eqr{EQddft}
on the continuity equation,
\begin{align}
  \frac{\partial \rho(\rv,t)}{\partial t}=  - \nabla \cdot \Jv(\rv,t),
  \label{EQcontinuity}
\end{align}
where $\Jv\rt$ is the microscopically resolved one-body current
distribution. Equation \eqref{EQcontinuity} is exact in a variety of
contexts, including overdamped Brownian dynamics, as described either
on the Fokker-Planck level by the Smoluchowski equation or by the
corresponding overdamped Langevin equation that governs the
trajectories, as they are realized in simulation work
\cite{sammueller2021}.  For BD the one-body current distribution is
given exactly by \cite{schmidt2022rmp}:
\begin{align}
  \gamma \Jv(\rv,t) &= -k_BT \nabla\rho(\rv,t)
  + \Fv_{\rm int}(\rv,t) + \rho(\rv,t) \fv_{\rm ext}(\rv,t).
  \label{EQcurrentBD}
\end{align}
This identity expresses the force density balance of the negative
friction force density (left hand side) with the force densities due
to ideal thermal diffusion, interparticle interactions, and external
influence (three contributions on the right hand side). Here the
interparticle force density distribution is given by the statistical
average
\begin{align}
  \Fv_{\rm int}(\rv,t) &= -\Big\langle
  \sum_i \delta(\rv-\rv_i)\nabla_i u(\rv^N)
  \Big\rangle\Big|_t,
  \label{EQFintAverage}
\end{align}
where the angular brackets indicate an average at fixed time~$t$ over
the nonequilibrium many-body distribution, $u(\rv^N)$ is the
interparticle interaction potential that depends on all particle
position coordinates $\rv^N\equiv \rv_1,\ldots,\rv_N$ and $\nabla_i$
indicates the derivative with respect to the position $\rv_i$ of
particle $i$.  The formulation of \eqr{EQFintAverage} is based on the
concept of static operators and a dynamically evolving probability
distribution. This is analogous to the Schr\"odinger picture of
quantum mechanics. The Heisenberg picture is more closely related to
simulation work. Here the probability distribution is that of the
initial microstates and the operators move forward in time, i.e., the
position $\rv_i(t)$ of particle $i$ changes over the course of
time. Then the Dirac distribution in \eqr{EQFintAverage} becomes
$\delta(\rv-\rv_i(t))$, with the generic position variable~$\rv$
however remaining static. The forces are those that act in the given
microstate $\rv^N(t)$ at time $t$, i.e., the interparticle force on
particle $i$ at time $t$ is $-\nabla_i u(\rv^N(t))$.

In practice, using BD simulations, carrying out the average in
\eqr{EQFintAverage} requires to build the mean over sufficiently many
separate realizations of the microscopic evolution of the many-body
system that differ {\ms in the initial microstate (as e.g.\ drawn from
  an equilibrium ensemble)} and in the realization of the thermal
noise.  As \eqr{EQFintAverage} measures both the probability to find
particle $i$ at position $\rv$ (via the delta function) and the
interparticle force that acts via the negative gradient $-\nabla_i
u(\rv^N)$, we refer to $\Fv_{\rm int}\rt$ as a {\it force
  density}. The corresponding {\it force field} $\fv_{\rm int}\rt$ is
obtained by simple normalization with the density profile,
i.e.\ $\fv_{\rm int}\rt=\Fv_{\rm int}\rt/\rho\rt$. Building this ratio
scales out the probability effect and the force field then carries
physical units of force, i.e.\ energy per length.

In equilibrium the definition \eqref{EQFintAverage} remains intact.
Complementing the statistical average, static density functional
theory allows to express the equilibrium force density as being
functionally dependent on the density profile via the functional
derivative of the excess free energy functional according to:
\begin{align}
  \Fv_{\rm int}(\rv)\big |_{\rm eq} &= -\rho(\rv) \nabla 
  \frac{\delta F_{\rm exc}[\rho]}{\delta\rho(\rv)}.
  \label{EQFintEqFunctional}
\end{align}
Crucially, and in contrast to \eqr{EQFintAverage}, here the internal
force density is directly expressed as a density functional. This
dependence has superseded the original dependence on the external
potential, as is manifest in the probability distribution for building
the average~\eqref{EQFintAverage} in equilibrium.

As a self-consistency check we insert the force density functional
\eqref{EQFintEqFunctional} into the equilibrium limit of the force
density balance \eqref{EQcurrentBD}. The current vanishes in the
equilibrium case, $\Jv\rt\equiv 0$, and we obtain
\begin{align}
  -k_BT \nabla\rho(\rv) + \Fv_{\rm int}(\rv)|_{\rm eq}
  + \rho(\rv) \fv_\ext(\rv) &= 0.
  \label{EQybg}
\end{align}
This result is independent of time and it constitutes the gradient of
the static Euler-Lagrange equation \eqref{EQdft} when divided by the
density profile. (Insert \eqr{EQFintEqFunctional}, identify the ideal
gas contribution $-k_BT\nabla\rho(\rv)=-\rho(\rv)\delta F_{\rm
  id}[\rho]/\delta\rho(\rv)$, and divide by $\rho(\rv)$.) The
classical force density balance result \eqref{EQybg} by Yvon, Born and
Green \cite{hansen2013} has recently been derived from systematically
addressing thermal Noether invariance
\cite{hermann2021noether,hermann2022topicalReview} against locally
resolved spatial deformations of the statistical ensemble
\cite{tschopp2022forceDFT,sammueller2022forceDFT,hermann2022quantum},
as also valid quantum mechanically~\cite{hermann2022quantum} and at
second order in the displacement field
\cite{hermann2022variance,sammueller2023whatIsLiquid}; we give a brief
account of this theory in Sec.~\ref{SECnoether} below.

A naive transfer of \eqr{EQFintEqFunctional} to nonequilibrium lets
one simply evaluate the equilibrium excess free energy functional at
the instantaneous nonequilibrium density $\rho\rt$. In order to
separate this contribution from true static equilibrium, we refer to
this force density as being adiabatic (subscript ``ad'') and to be
defined as
\begin{align}
  \Fv_{\rm ad}\rt &= -\rho\rt 
  \nabla \frac{\delta F_{\rm exc}[\rho]}{\delta\rho\rt}.
  \label{EQFad}
\end{align}
We recall that the right hand side offers a concrete computational
structure that is of practical usefulness in actual applications, as
considerable knowledge about approximative forms of the excess free
energy density functional $F_{\rm exc}[\rho]$ is available.  Using the
adiabatic force density as a proxy for the true nonequilibrium
intrinsic force density distribution~\eqref{EQFintAverage},
i.e.\ setting $\Fv_{\rm int}\rt=\Fv_{\rm ad}\rt$ in the force density
balance \eqref{EQcurrentBD} together with the continuity equation
\eqref{EQcontinuity} leads to the DDFT equation of
motion~\eqref{EQddft}. The adiabatic force density approximation is
uncontrolled though and the theory inherently yields the dynamics as
an adiabatic sequence of equilibrium states. Surely, more than 40
years after the conception of the DDFT \cite{evans1979}, we have to be
able to do better!

\section{Power functional techniques}
\label{SECpft}

Power functional theory \cite{schmidt2022rmp} offers a concrete
mathematical structure to go forward. We describe the essential steps
that enable one to go beyond the DDFT and to hence address a
significantly expanded realm of nonequilibrium physics which
\eqr{EQddft} is oblivious of.

The interparticle force density profile \eqref{EQFintAverage} is
identified to consist of two contributions according to:
\begin{align}
  \Fv_{\rm int}\rt &= \Fv_{\rm ad}\rt + \Fv_{\rm sup}\rt.
  \label{EQFintSplitting}
\end{align}
Here $\Fv_{\rm ad}\rt$ is the adiabatic force density profile, as
given formally via the explicit equilibrium free energy derivative
\eqref{EQFad} and directly accessible in simulations via the custom
flow method \cite{delasheras2019customFlow,renner2021customFlowMD}.
The custom flow algorithm allows to systematically construct a
hypothetical adiabatic (equilibrium) system that shares its density
profile with the nonequilibrium system at the given time. Then
sampling the internal force density in the adiabatic system yields
results for $\Fv_\ad\rt$.

The second, superadiabatic contribution in \eqr{EQFintSplitting},
$\Fv_\sup\rt$, contains all effects that are not expressible as an
instantaneous density functional. This includes forces that lead to
viscous and to nonequilibrium structure forming phenomena, as we
exemplify below in a concrete model compressional flow situation.
Formally, the superadiabatic force density is generated from the
superadiabatic excess free power functional $P_t^{\rm exc}[\rho,\Jv]$
upon functional differentiation with respect to the one-body current
via \cite{schmidt2022rmp,schmidt2013pft}:
\begin{align}
  \Fv_{\rm sup}\rt &= 
  -\rho\rt\frac{\delta P_t^{\rm exc}[\rho,\Jv]}
  {\delta \Jv(\rv,t)}.
  \label{EQFsup}
\end{align}
The functional dependence of $P_t^{\rm exc}[\rho,\Jv]$ on the density
and current is causal, i.e.\ on the values of these fields at prior
times to $t$; density and current need to satisfy the continuity
equation. Upon using Eqs.~\eqref{EQFintSplitting} the force density
balance \eqref{EQcurrentBD} attains the following form:
\begin{align}
  \gamma \Jv\rt &=
  -k_BT \nabla\rho\rt + \Fv_{\rm ad}\rt 
  \notag\\&
  \qquad + \Fv_{\rm sup}\rt +\rho\rt \fv_{\rm ext}\rt.
  \label{EQforceDensityBalance}
\end{align}
This relationship holds beyond gradient forms of $\fv_\ext\rt$,
i.e.\ for external force fields that contain non-conservative
contributions. Crucially, $\Fv_\sup\rt$ will in general also acquire
nonconservative contributions, such as e.g.\ damping effects that
represent viscous behaviour. Moreover, nonequilibrium
structure-forming effects will also arise in general. These affect
directly the shape of the density profile, whether this evolves in
time or persists in a nonequilibrium steady state.

If one wishes to eliminate the explicit occurrence of the current from
the dynamics, then inputting the force density balance
\eqref{EQforceDensityBalance} into the continuity equation
\eqref{EQcontinuity} leads to the following formally exact form of the
equation of motion for the density profile:
\begin{align}
  \frac{\partial \rho\rt}{\partial t} &=
  D_0 \nabla^2\rho\rt
  + \nabla\cdot\frac{\rho\rt}{\gamma} 
  \nabla \frac{\delta F_{\rm exc}[\rho]}{\delta\rho\rt}
  \notag\\& \qquad
  -\nabla\cdot\frac{\rho\rt}{\gamma}
  [\fv_{\rm sup}\rt + \fv_\ext\rt].
  \label{EQofMotionForDensity}
\end{align}
Here it is apparent that the superadiabatic force field
$\fv_\sup\rt=\Fv_\sup\rt/\rho\rt$ has a direct effect on the system
dynamics. The effect is similar to that of the external force
field. Crucially though, both force fields are independent of each
other: the external force field represents a prescribed and inert
influence on the system.  In contrast, the superadiabatic force field
is an emergent phenomenon that arises due to interparticle
interactions and, from the functional point of view, depends
non-locally in position and causally in time on the one-body density
and on the current profile.

\begin{figure*}
  \includegraphics[width=0.51\linewidth]{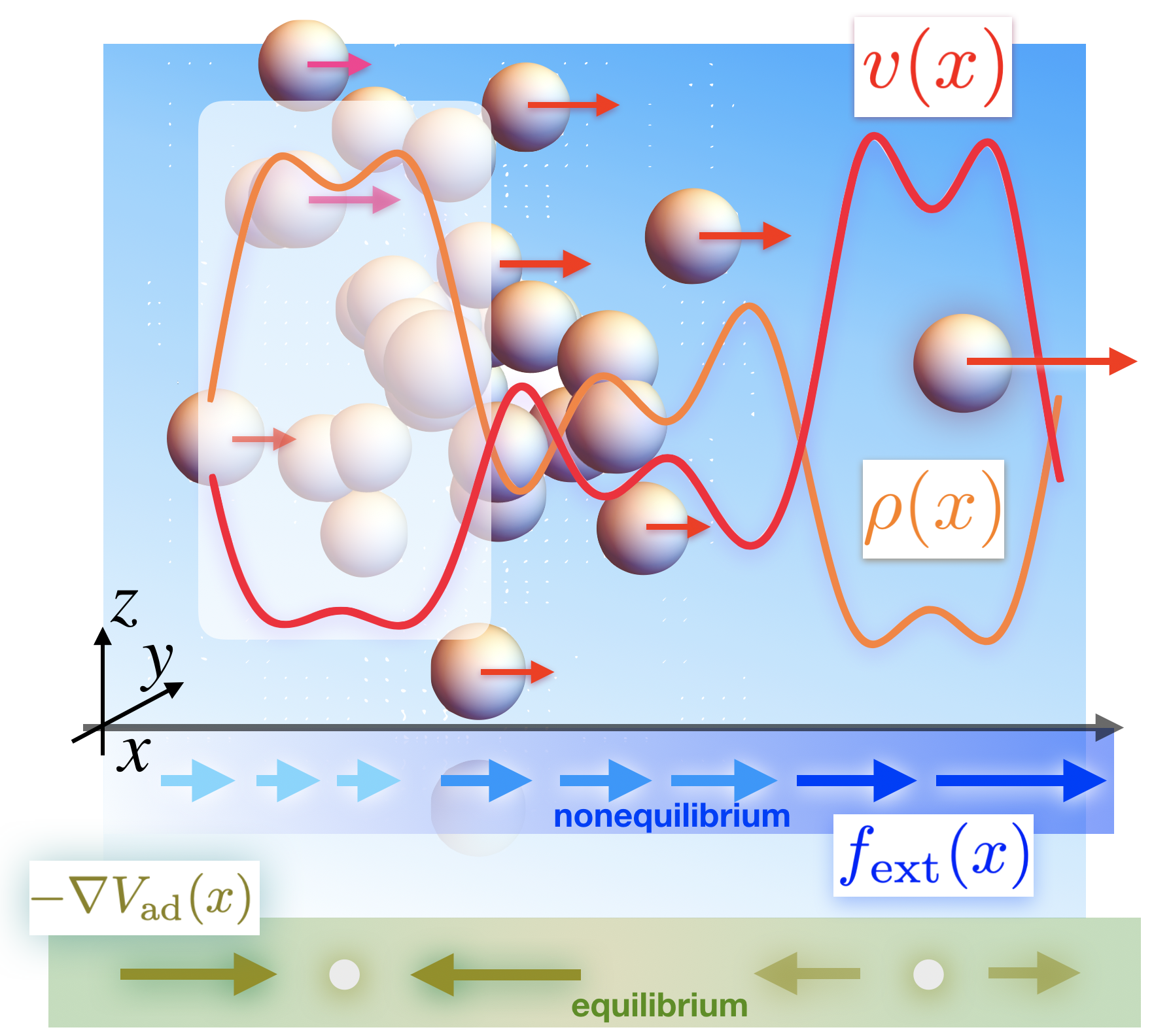}\\
  \caption{Illustration of unidirectional compressional flow of a
    liquid. The three-dimensional system is set into motion (red
    arrows) by the action of an external force profile $f_\ext(x)$
    (blue arrows) which acts along the $x$-axis. The system retains
    planar geometry such that spatial inhomogeneities only occur as a
    function of $x$. The density profile $\rho(x)$ (orange curve) and
    the velocity profile $v(x)$ (red curve) are both stationary in
    time but inhomogeneous in position. The local one-body current
    $J(x)=\rho(x)v(x)=\rm const$ and as a result the system is in a
    nonequilibrium steady state. The corresponding adiabatic system is
    in equilibrium (it has no mean flow) and it has by construction an
    unchanged density profile $\rho(x)$.  In the adiabatic system the
    spatial variation of $\rho(x)$ is stabilized by the action of an
    external force field $-\nabla V_\ad(x)$ (olive arrows), which acts
    solely in the adiabatic system.
    \label{FIGillustration}
  }
\end{figure*}

Although setting $\fv_{\rm sup}\rt= 0$ yields the DDFT \eqref{EQddft},
the superadiabatic force field $\fv_\sup\rt$ was demonstrated to exist
\cite{fortini2014prl, stuhlmueller2018prl, treffenstaedt2020shear,
  jahreis2019shear, delasheras2018velocityGradient,
  delasheras2020fourForces, sammueller2022gel} and in general to play
a major role in the dynamics on the one-body level and, based on
test-particle concepts \cite{percus1962, archer2007dtpl,
  hopkins2010dtpl, stopper2015jcp, stopper2015pre, brader2015dtpl},
two-body correlation functions
\cite{schindler2016dynamicPairCorrelations, treffenstaedt2021dtpl,
  treffenstaedt2022dtpl}, and for active matter \cite{hermann2019prl,
  hermann2019pre, krinninger2016, krinninger2019, hermann2021molPhys}.
Both the flow properties as well as the spatial structure formation in
the system are affected.

To reveal additional physics, it is useful to split into
``structural'' and ``flow'' contributions. This was established
e.g.\ for complex flow patterns that occur in driven BD
\cite{stuhlmueller2018prl,delasheras2020fourForces}, for active
Brownian particles which form a self-sustained interface at
motility-induced phase coexistence \cite{hermann2019pre,
  hermann2019prl, hermann2021molPhys, krinninger2016, krinninger2019},
as well as very recently for a sheared three-body colloidal gel former
\cite{sammueller2022gel}. Before we demonstrate these concepts for an
example of steady nonequilibrium below, we first describe two simple
model power functionals that respectively generate structure and
viscously dampen the motion and that, as we will see, give a good
account of the nonequilibrium flow considered below.

We concentrate on the low-order terms that are relevant for
compressional/extensional flow, i.e., for situations where
$\nabla\cdot\vel\rt\neq 0$. We focus on cases where there is no
rotational motion (such as shearing) and hence
$\nabla\times\vel\rt=0$. The velocity gradient superadiabatic power
functional consists of a sum,
\begin{align}
  P_t^\exc[\rho,\vel]
  &= P_t^\flow[\rho,\vel] + P_t^\str[\rho,\vel].
  \label{EQPtexcSplitting}
\end{align}
Here the flow and structural \cite{stuhlmueller2018prl,
  delasheras2020fourForces} contributions are approximated,
respectively, by the following time-local (Markovian) and
space-semilocal (i.e.\ involving $\nabla$) forms
\begin{align}
  P_t^\flow[\rho,\vel] &= \frac{\eta}{2} \int d\rv
  [\rho\rt\nabla\cdot\vel\rt]^2,
  \label{EQPtexcFlow}
  \\
  P_t^\str[\rho,\vel] &= -
  \frac{\chi}{3} \int d\rv  [\rho\rt\nabla\cdot\vel(\rv,t)]^3,
  \label{EQPtexcStructure}
\end{align}
where the overall prefactors $\eta$ and $\chi$ control the respective
magnitude {\ms and they play the role of transport coefficients (see
  below).}  The flow functional \eqref{EQPtexcFlow} is quadratic both
in density and in the velocity field; the structural
functional~\eqref{EQPtexcStructure} is of cubic order in each of these
variables. Explicit higher-order functionals exist
\cite{delasheras2020fourForces} and they become relevant when driving
the system strongly.  We will return to the consequences of
Eqs.~\eqref{EQPtexcFlow} and \eqref{EQPtexcStructure} after laying out
in Sec.~\ref{SECsteadyStates} the actual flow situation that we use as
a model to exemplify the implications for the physics. Before doing
so, we briefly describe several further key aspects of the power
functional framework.

Power functional theory provides a formal framework for the inclusion
of time- and space-nonlocal dynamics \cite{renner2022prl,
  treffenstaedt2021dtpl, treffenstaedt2020shear}.  While \eqr{EQFsup}
applies to overdamped dynamics, the acceleration field becomes a
further relevant degree of freedom if inertia are relevant
\cite{schmidt2018md, renner2022prl, schmidt2015qpft,
  bruetting2019viscosity} whether classically in molecular dynamics
\cite{schmidt2018md,renner2022prl} or in quantum dynamics
\cite{schmidt2015qpft, bruetting2019viscosity}.  Here the memory
functions act as convolution kernels on specific kinematic fields and
rotational and compressional contributions to the dynamics are
genuinely built in.  As laid out above, the framework is based on an
exact variational concept \cite{schmidt2013pft,schmidt2022rmp}, and
the resulting functional mapping was shown to be explicitly accessible
in many-body simulation via the custom flow computer simulation
method~\cite{delasheras2019customFlow, renner2021customFlowMD}.

Even simple mathematical model forms for the nonequilibrium
contribution to the power functional, such as Eqs.~\eqref{EQPtexcFlow}
and \eqref{EQPtexcStructure}, already capture essential physics (as we
demonstrate below) and dynamical two-body correlation functions are
accessible via test particle dynamics \cite{percus1962,
  archer2007dtpl, hopkins2010dtpl, brader2015dtpl,
  schindler2016dynamicPairCorrelations, treffenstaedt2021dtpl,
  treffenstaedt2022dtpl, bier2008prl, stopper2015jcp, stopper2015pre,
  stopper2018dtpl}.  The power functional is thereby not to be
confused with the often vague concept of a ``nonequilibrium free
energy''. The proper equilibrium free energy functional does play a
central role in power functional theory though, via providing the
description of the adiabatic reference state \cite{schmidt2022rmp},
see the generation of the force density distribution via
  functional differentiation \eqref{EQFad}, as is relevant for the
  interparticle force splitting~\eqref{EQFintSplitting}, and the full
  density equation of motion \eqref{EQofMotionForDensity}.

The relevance of superadiabatic contributions to the dynamics,
i.e.\ of those effects that lie beyond \eqr{EQddft}, has been amply
demonstrated in the literature
\cite{fortini2014prl,schindler2016dynamicPairCorrelations,
  delasheras2018velocityGradient,
  stuhlmueller2018prl,jahreis2019shear, delasheras2020fourForces,
  treffenstaedt2020shear,treffenstaedt2021dtpl,treffenstaedt2022dtpl}.
Both adiabatic and superadiabatic effects arise from integrating out
the dynamical degrees of freedom of the many-body problem.

Ensemble differences between canonical dynamics and grand canonical
equilibrium have been systematically
addressed~\cite{delasheras2014canonical,
  delasheras2016particleConserving, schindler2019particleConserving}
and these do not account for the observed differences between
adiabatic and superadiabatic dynamics.  The kinematic dependence on
the motion of the system arises formally \cite{schmidt2022rmp}, it can
be explicitly traced in many-body computer simulation work
\cite{delasheras2020fourForces}, and it is amenable to machine
learning, as we demonstrate in Sec.~\ref{SECmachineLearning}. Before
doing so, we first formulate the representative flow problem that we
will use to apply the above concepts.

\section{Nonequilibrium steady states}
\label{SECsteadyStates}
We restrict ourselves to flow situations with one-body fields that are
inhomogeneous in position but independent of time, i.e.\ $\rho(\rv)$
and $\vel(\rv)$. Then trivially $\partial\rho(\rv)/\partial t=0$ and
the continuity equation \eqref{EQcontinuity} constrains both fields to
satisfy $\nabla\cdot[\rho(\rv)\vel(\rv)]=0$. As a representative case
we illustrate in Fig.~\ref{FIGillustration} a nonequilibrium steady
state of a three-dimensional liquid undergoing unidirectional
compressional flow. Flow along a single given direction occurs
e.g.\ under the influence of gravity, where sedimentation of colloids
leads to both compression in the lower parts of the sample and
expansion in the upper parts of the sample. Here we disregard
transient phenomena and investigate an idealized periodic system,
where flowing steady states can form.

{\ms This chosen uniaxial flow in planar geometry is complimentary to
  DDFT, as density gradients are relevant and the density profile
  alone already contains much non-trivial information about the
  dynamics that the system undergoes. Hence this specific geometry is
  often used to carry out generic tests; see e.g.\ the investigation
  of the quality of force-based DDFT~\cite{tschopp2022forceDFT,
    sammueller2022forceDFT}.
In contrast, shear flow is very different, as any motion that occurs
perpendicular to the density gradient is not captured by
Eq.~\eqref{EQddft}; we refer the reader to
Refs.~\cite{tevrugt2020review, tevrugt2022perspective} for a
description of efforts to include these effects within DDFT via
different types of modifications of~Eq.~\eqref{EQddft}.

}

In order to elucidate the physics in the chosen uniaxial compressional
setups, we follow the splitting \eqref{EQPtexcSplitting} of the
superadiabatic power functional into structural and flow contributions
and hence decompose the superadiabatic force field accordingly as
\begin{align}
  \fv_\sup(\rv)=\fv_\str(\rv) + \fv_\flow(\rv),
  \label{EQflowStructureSplitting}
\end{align}
where the right hand side consists of the nonequilibrium structural
force field $\fv_\str(\rv)$ and the flow force field~$\fv_\flow(\rv)$.
Both of these force contributions arise from the microscopic
interparticle interactions, as coarse-grained in a microscopically
sharp way to the one-body level. We lay out in the following the
benefits of the structure-flow splitting
\eqref{EQflowStructureSplitting} and its definition via flow reversal
symmetry.

First, on the more practical level, \eqr{EQflowStructureSplitting}
allows to carry out a corresponding splitting of the force density
balance~\eqref{EQforceDensityBalance} [we divide by $\rho(\rv)$ to
  obtain force fields]. The result is a set of two coupled equations
of motion, with one of them depending explicitly on the velocity
profile and the second one depending explicitly on the density
profile:
\begin{align}
  \gamma\vel(\rv) &=\fv_\flow(\rv) + \fv_{\ext,\rm f}(\rv),
  \label{EQforceBalanceFlow}\\
  0 &= 
  \fv_\str(\rv)  
  -k_BT \nabla \ln \rho(\rv) + \fv_\ad(\rv) 
  +\fv_{\ext,\rm s}(\rv).
  \label{EQforceBalanceStructure}
\end{align}
Building the sum of Eqs.~\eqref{EQforceBalanceFlow} and
\eqref{EQforceBalanceStructure} and multiplying by the density profile
restores the full force density balance
\eqref{EQforceDensityBalance}. The external force field is split
according to $\fv_\ext(\rv) = \fv_{\ext,\rm f}(\rv) +\fv_{\ext, \rm
  s}(\rv)$, where the two terms couple to the flow via $\fv_{\ext,\rm
  f}(\rv)$ in \eqr{EQforceBalanceFlow} and to the structure via
$\fv_{\ext,\rm s}(\rv)$ in \eqr{EQforceBalanceStructure}.

On the superficial level the two equations \eqref{EQforceBalanceFlow}
and \eqref{EQforceBalanceStructure} appear to be independent of each
other, as no single field appears explicitly in both equations.
However, the two equations are indeed intimately coupled to each other
by the interparticle interactions, as represented by both the
adiabatic and the two superadiabatic (flow and structural) force
fields. These three intrinsic force contributions provide the physical
representation of the true nonequilibrium steady state dynamics.

The flow-structure splitting \eqref{EQflowStructureSplitting} is
uniquely determined by the symmetry properties of the forces upon
motion reversal of the system \cite{delasheras2020fourForces}. Motion
reversal is a discrete symmetry operation, and hence different from
continuous invariances where Noether's theorem applies
\cite{hermann2021noether, hermann2022topicalReview,
  hermann2022quantum, tschopp2022forceDFT, sammueller2022forceDFT,
  hermann2022variance, sammueller2023whatIsLiquid}. One considers a
``reversed'' system, which is also in steady state and possesses an
unchanged density profile $\rho(\rv)$. The flow, however, is directed
against the velocity orientation in the original ``forward''
system. Hence the velocity profile in the reversed system is simply
$-\vel(\rv)$. As a result the current also acquires a minus sign,
$-\rho(\rv)\vel(\rv)$, which however does not affect the (vanishing)
divergence, $\nabla\cdot[-\rho(\rv)\vel(\rv)]=0$. Thus the reversed
state indeed is stationary. The two superadiabatic contributions are
then defined to be unchanged [$\fv_\str(\rv)$] and inverted
[$-\fv_\flow(\rv)$] in the reversed system. Consequentially, the
superadiabatic force field in the reversed system is the difference
$\fv_\str(\rv)-\fv_\flow(\rv)$.

Analyzing the symmetry properties of the adiabatic force field is
straightforward. We recall that $\fv_\ad(\rv)$ is a density functional
via \eqr{EQFad}. The density profiles in the forward and in the
reversed systems are identical though. Hence $\fv_\ad(\rv)$ is
invariant under motion reversal. Motion reversal is a useful device in
order to i) rationalize the nonequilibrium behaviour according to the
split force balance \eqref{EQforceBalanceFlow} and
\eqref{EQforceBalanceStructure}, and to ii) classify the dependence of
superadiabatic forces on the velocity field into even powers, which
constitute $\fv_\str(\rv)$, and odd powers, which form
$\fv_\flow(\rv)$.

We can demonstrate this mechanism explicitly on the basis of the above
flow and structural power functionals \eqref{EQPtexcFlow} and
\eqref{EQPtexcStructure}. Superadiabatic force fields are generated
via the functional derivative \eqref{EQFsup} with respect to the
current or, analogously, by functionally deriving by $\vel\rt$ and
dividing the result by $\rho\rt$.  The resulting superadiabatic
one-body force field consists of two components. The viscous flow
force and \cite{delasheras2018velocityGradient, stuhlmueller2018prl}
and the structural force follow respectively as
\begin{align}
  \fv_{\rm flow}(\rv) &=
  \frac{\eta}{\rho(\rv)}
  \nabla [\rho(\rv)^2\nabla\cdot\vel(\rv)],
  \label{EQsupForceFlow}
  \\
  \fv_{\rm str}(\rv) &=  -
  \frac{\chi}{\rho(\rv)} \nabla \{\rho(\rv)^3
      [\nabla \cdot \vel(\rv)]^2\},
  \label{EQsupForceStructural}
\end{align}
where Eq.~\eqref{EQsupForceFlow} is odd (linear) and
Eq.~\eqref{EQsupForceStructural} is even (quadratic) in the
derivatives of the velocity field, as desired {\ms and we re-iterate
  that both expressions are only valid for small enough velocity
  gradients.}

One might wonder where all this genuine nonequilibrium physics leaves
the DDFT!  Some readers will find the instantaneous dynamics, as
generated from an adiabatic free energy according to \eqref{EQddft},
to be more appealing and intuitive than the thinking in terms of the
above described apparently intricate functional relationships. Why not
live with \eqr{EQddft}, use it, and simply accept its defects?  In
order to address this question and to demonstrate why this path is
severely restricted from the outset, we turn in
Sec.~\ref{SECmachineLearning} to an explicit demonstration of the
functional relationship that governs the nonequilibrium physics,
i.e.\ the kinematic functional map from the one-body mean motion to
the internal force field. Before doing so, we demonstrate that
Noether's theorem of invariant variations has much to say about our
present setup.

\begin{figure*}
  \includegraphics[width=0.99\linewidth]{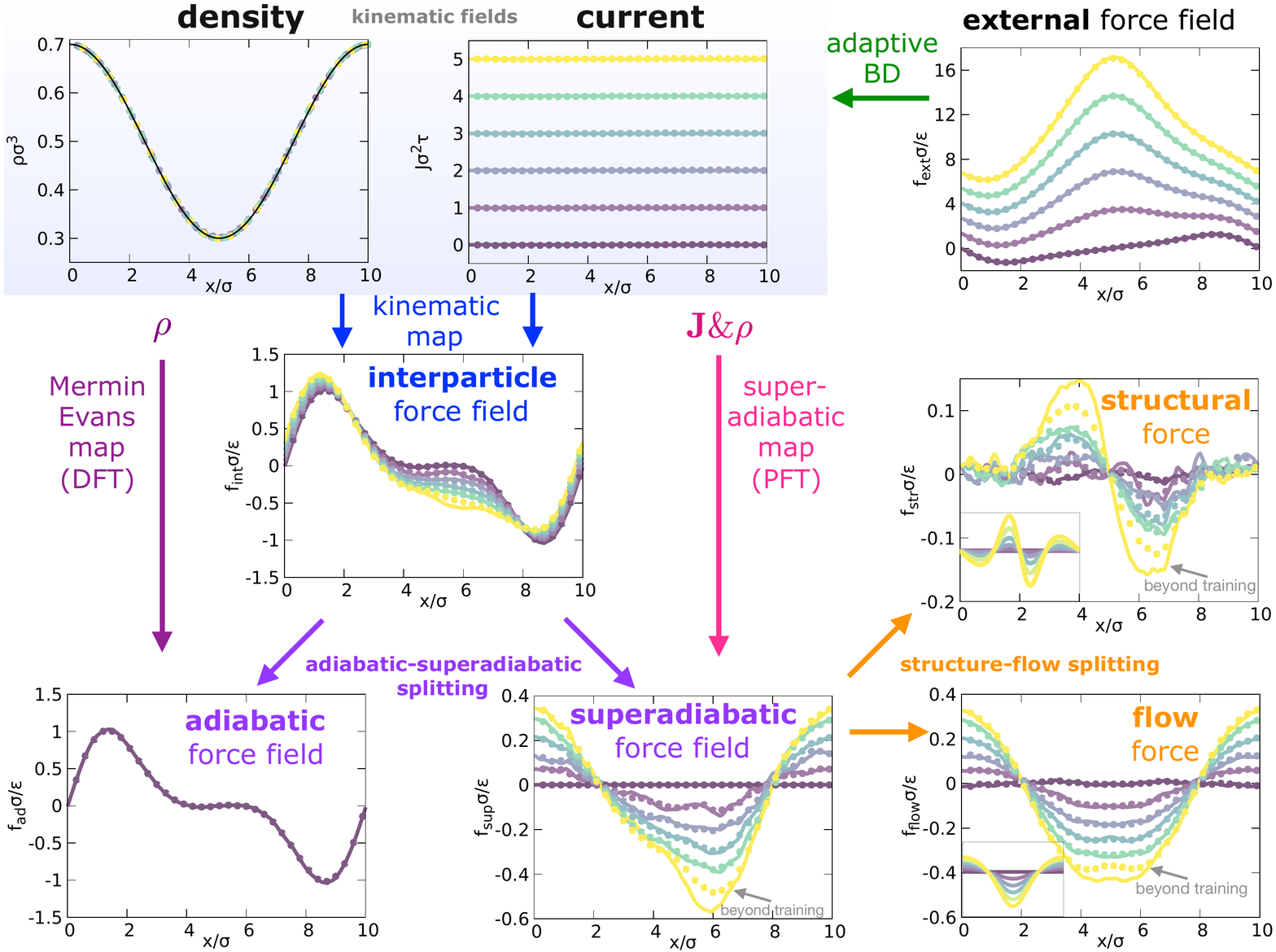}
  \caption{Kinematic profiles and force fields for uniaxial
    compressional flow of the LJ fluid.  Results are shown from
    machine learning (lines) and from direct adaptive BD simulations
    (symbols). Functional relationships are represented by vertical
    arrows.  Shown are the density profile $\rho(x)$, the one-body
    current $J(x)$ and the external force field $f_\ext(x)$ (top row)
    as a function of the scaled distance $x/\sigma$, where $\sigma$ is
    the LJ length scale {\ms and $k_BT/\epsilon=1.5$ throughout}.  The
    density and the current functionally determine both the
    interparticle force field $f_{\rm int}(x)$ via the kinematic map
    and the superadiabatic force field $f_\sup(x)$ via the
    superadiabatic kinematic map (middle row).  The internal force
    field $f_{\rm int}(x)$ splits into superadiabatic and adiabatic
    force contributions.  The adiabatic force field $f_\ad(x)$ is a
    density functional via the Mermin-Evans map of density functional
    theory. The structural and flow force fields are split according
    to their symmetry upon motion reversal.  The colour code
    represents different values of the current
    $J_0\sigma^2\tau=0,1,2,3,4,5$ (from violet to yellow, see the
    center panel in the top row). The two insets show the predictions
    from the analytical velocity gradient functionals
    \eqref{EQsupForceFlow} and \eqref{EQsupForceStructural} {\ms on
      the same scale as the respective main panel; the transport
      coefficients are chosen as $\eta/(\eps \tau\sigma^3 )=0.35$ and
      $\chi/(\eps \tau^2 \sigma^6 ) =0.075$ to give good agreement
      with the quasi-exact data. } The system with $J_0=0$ is at rest
    in equilibrium and it doubles as the adiabatic state because its
    density profile is identical to that of the flowing systems (as
    shown in the first panel).
    {\ms The small differences in superadiabatic forces from BD and
      from machine learning for the case of the highest current
      considered, $J_0\sigma^2\tau=5$, occur as this value is beyond
      those encountered in the training data.}
    \label{FIGprofiles}
  }
\end{figure*}

\section{Noether force sum rules}
\label{SECnoether}

We discuss one of the arguably simplest cases of exploitation of the
inherent symmetries of a thermal many-body system, that of global
translational invariance of its statistical mechanics
\cite{hermann2021noether,hermann2022topicalReview}.  We consider a
``shifting'' transformation, where all particle coordinates change
according to the map $\rv_i\to\rv_i + \eps$, where $\eps=\rm const$.
This uniform shifting operation leaves all interparticle distance
unchanged, $\rv_i-\rv_j\to(\rv_i+\eps)-(\rv_j+\eps)\equiv\rv_i-\rv_j$.
As a consequence the interparticle potential is invariant under the
transformation, which we can express as the identity
$u(\rv_1,\ldots,\rv_N)=u(\rv_1+\eps,\ldots,\rv_N+\eps)$. Here equality
holds irrespectively of the magnitude and the direction of the
shifting vector $\eps$.

The Noether argument proceeds with a twist. Despite the absence of
dependence on $\eps$, we can nevertheless differentiate both sides of
the equation with respect to $\eps$ and the result will be a valid
identity. We obtain $0=\partial
u(\rv_i+\eps,\ldots,\rv_N+\eps)/\partial \eps = \sum_i \nabla_i
u(\rv_1,\ldots,\rv_N)$, where we have set $\eps=0$ after taking the
derivative. We multiply by $-1$ and insert $1=\int
d\rv\delta(\rv-\rv_i)$, which yields
\begin{align}
  -\int d\rv \sum_i \delta(\rv-\rv_i) \nabla_i u(\rv^N) &= 0.
  \label{EQFintOperatorZero}
\end{align}
The expression on the left hand side allows to identify the locally
resolved interparticle force operator $\hat\Fv_{\rm int}(\rv)=-\sum_i
\delta(\rv-\rv_i)\nabla_i u(\rv^N)$, such that
\eqr{EQFintOperatorZero} attains the form $\int d\rv \hat\Fv_{\rm
  int}(\rv)=0$. This identity holds for each microstate $\rv^N$ and
hence it remains trivially valid upon averaging over the many-body
distribution function, irrespective of whether this is in- or
out-of-equilibrium. We can hence conclude the vanishing of the global
interparticle force, expressed as the integral over the mean force
density $\Fv_{\rm int}(\rv)=\langle \hat\Fv_{\rm int}(\rv)\rangle$ as
\begin{align}
  \int d\rv \Fv_{\rm int}(\rv,t) &= 0.
  \label{EQsumRuleFint}
\end{align}
Equation \eqref{EQsumRuleFint} holds at all times $t$ and it can be
viewed as a consequence of Newton's third law, see the discussion in
Ref.~\cite{hermann2021noether}. Using the adiabatic-superadiabatic
force splitting \eqref{EQFintSplitting} one can further conclude that
the both global contributions need to vanish individually,
\begin{align}
  \int d\rv \Fv_\ad(\rv,t) &= 0,
  \label{EQsumRuleFad}\\
  \int d\rv \Fv_\sup(\rv,t) &= 0.
  \label{EQsumRuleFsup}
\end{align}
The proof can either be based on the fact that \eqr{EQsumRuleFad} is
merely \eqr{EQsumRuleFint} for the special case of an equilibrium
system, from which then \eqr{EQsumRuleFsup} follows from the force
splitting \eqref{EQFintSplitting}. Alternatively and starting from a
very fundamental point of view, the global translational invariance of
the excess free energy functional $F_{\rm exc}[\rho]$ and of the
superadiabatic free power functional $P_t^{\rm exc}[\rho,\vel]$, here
considered instantaneously at time $t$, lead directly to
Eqs.~\eqref{EQsumRuleFad} and \eqref{EQsumRuleFsup}, see
Refs.~\cite{hermann2021noether,hermann2022topicalReview} for detailed
derivations.

It is interesting to apply the Noether concept to the flow-structure
splitting \eqr{EQflowStructureSplitting} of the superadiabatic force
field.  One can see straightforwardly, from the symmetry upon motion
reversal, that both the global structural force and the global flow
force need to vanish individually:
\begin{align}
  \int d\rv\rho(\rv)\fv_\flow(\rv) &= 0,
  \label{EQsumRuleFflow}\\
  \int d\rv\rho(\rv)\fv_\str(\rv) &= 0.
  \label{EQsumRuleFstr}
\end{align}
We prove by contradiction and assume that it is not the case,
i.e.\ that each integral gives the same global force, but with
opposite sign, such that the sum vanishes and \eqr{EQsumRuleFsup}
remains valid. Per construction, $\fv_\flow(\rv)$ changes sign in the
motion reversed system, but $\fv_\str(\rv)$ does not. Hence
\eqr{EQsumRuleFsup} can only be satisfied in the motion-reversed
system provided that both the flow and structural contribution vanish
separately.

We can explicitly test the validity of the sum
rules~\eqref{EQsumRuleFflow} and \eqref{EQsumRuleFstr} for the above
analytical force approximations \eqref{EQsupForceFlow} and
\eqref{EQsupForceStructural}. The respective integrals are $\eta \int
d\rv \nabla[\rho(\rv)^2\nabla\cdot\vel(\rv)]=0$ and $\chi\int d\rv
\nabla \{\rho(\rv)^3[\nabla\cdot\vel(\rv)]^2\}=0$, which follows from
the divergence theorem, as boundary terms vanish. Hence the simple
non-local velocity gradient power functional approximations
\eqref{EQPtexcFlow} and \eqref{EQPtexcStructure} have passed the
global Noether validation test. This is nontrivial, as the proof rests
on the specific structure of the integrands being gradients, which for
more general analytical forms will not be the case. This exemplifies
the merits of Noether sum rules for assessing and by extension also
constructing theoretical nonequilibrium force approximations.

The Noether concept carries much further. {\ms Reference
  \cite{hermann2021noether} presents a generalization of the global
  sum rules, such as the vanishing of the total superadiabatic force
  \eqref{EQsumRuleFsup}, for so-called time direct correlation
  functions. These are defined via functional derivatives of the
  superadiabatic power functional, in generalization of the
  superadiabatic force density as generated via the derivative
  \eqref{EQFsup} with respect to the current distribution. We have
  shown \cite{hermann2021noether} that these time direct correlation
  functions satisfy additional memory sum rules and we expect} the
corresponding identities to be helpful in the study of temporal
nonlocality.  Further work was addressed at the variance of global
fluctuations, which were shown to be constrained by Noether invariance
at the second order global level \cite{hermann2022variance}. Noether's
theorem also yields the locally resolved force balance relationship in
quantum mechanical many-body systems \cite{hermann2022quantum}.  Very
recently, striking two-body force-force and force-gradient correlation
functions for the precise and novel characterization of disordered
(liquid and gel) systems \cite{sammueller2023whatIsLiquid} were
revealed.  Exploiting Noether's concept in a statistical mechanical
setting is robust against changes of ensemble,
Ref.~\cite{hermann2022topicalReview} presents the transfer of the
grand ensemble formalism \cite{hermann2021noether} to canonical
systems. Considering global rotational invariance leads to (classical)
spin-orbit coupling of torque identities \cite{hermann2021noether}.

We return to steady states and demonstrate that the seemingly entirely
formal functional relationships do in fact apply to real systems. We
present in the following new computational methodology that we use to
demonstrate the functional point of view.  We will also demonstrate
that the sum rules \eqref{EQsumRuleFsup} and \eqref{EQsumRuleFflow}
are highly valuable in providing checks for numerical results.

\section{Machine learning the kinematic map}
\label{SECmachineLearning}

Machine learning proves itself to be an increasingly useful tool in a
variety of settings in soft matter, ranging from soft matter
characterization \cite{clegg2021ml}, engineering of colloidal
self-assembly \cite{dijkstra2021ml}, to the inverse design of soft
materials \cite{coli2022scienceAdvances}.  Pivotal studies were
addressed at colloidal structure detection \cite{boattinia2019ml}, the
identification of combinatorial rules in mechanical metamaterials
\cite{mastrigt2022prl}, the learning of many-body interaction
potentials for spherical \cite{campos2021ml} and for anisotropic
particles \cite{campos2022ml}, and the prediction of the dynamics of
supercooled liquids from their static properties~\cite{ciarella2022}.

{\ms
Concerning slow dynamics, machine learning was used for obtaining
memory kernels for generalised Langevin dynamics \cite{winter2023ml},
classifying the age \cite{janzen2023}, assessing the structural
heterogeneity \cite{paret2020}, and investigating dimensionality
reduction of local structure \cite{coslovich2022} of glasses.  Machine
learning was applied to equilibrium reactive processes such as
molecular isomerization \cite{singh2023ml} and to the behaviour of
rare diffusive molecular dynamics trajectories \cite{das2021}.

Machine learning plays an important role in the inverse design for
self-assembly of soft materials \cite{lindquist2016, shermann2020}.
Examples thereof include sequence-specific aggregation of copolymers
\cite{statt2021}, inverse design of multicomponent colloidal crystals
by reverse engineering the Hamiltonian of the system
\cite{mahynski2020}, characterizing the self-assembly of
three-dimensional colloidal systems \cite{oleary2021}, controlling
colloidal crystals via morphing energy landscapes \cite{zhang2020},
and learning free energy landscapes using artificial neural networks
\cite{sidky2018}.

In a liquid state theory-informed approach, Limmer and his coworkers
have considered potentials based on local representations of atomic
environments, in order to learn intermolecular forces at liquid-vapor
interfaces \cite{niblett2021}. They relate their machine-learning
approach to the local molecular field theory by Weeks and coworkers
\cite{weeks1995,weeks2002}, see Ref.\ \cite{archer2013lmft} for a
description of the relationship of this approach to DFT.

}

More specifically, in the context of classical density functional
theory, an early and pioneering study formulated a neural-network
approach to liquid crystal ordering in confinement
\cite{teixera2014}. Free energy density functionals were obtained for
one-dimensional fluids from a convolutional neural
network~\cite{lin2019ml} and an analytical form of an excess free
energy functional was generated from an equation learning
network~\cite{lin2020ml}. Cats et al.\ \cite{cats2022ml} recently used
machine learning to improve the standard mean-field approximation of
the excess Helmholtz free-energy functional for a three-dimensional
Lennard-Jones (LJ) system at a supercritical temperature. These
significant reserach efforts were devoted to tailoring analytical
forms of model free energy functionals, by training certain key
components such as spatial convolution kernels, and much insight into
the inner workings of excess free energy functionals was gained
\cite{lin2019ml, lin2020ml, cats2022ml}.
{\ms Very recent developments include using physics-constrained
  Bayesian inference of state functions \cite{yatsyshin2022} and to
  emulate functionals by active learning with error control
  \cite{fang2022}. The results of DFT calculations were also used as
  training data for investigating gas solubility in nanopores
  \cite{qiao2020}.

 }

However, here we proceed very differently and moreover do so
out-of-equilibrium.  We use the LJ model and the identical planar
geometry as in Ref.~\cite{cats2022ml}, such that the density profile
$\rho(x)$ depends only on a single position coordinate $x$. We
consider steady states and retain planar symmetry by considering flow
that is directed in the $x$-direction, such that the current
$\Jv(x)=J(x) \ev_x$, where $J(x)$ is the magnitude of the current and
$\ev_x$ is the unit vector in the $x$-direction. Both the density
profile $\rho(x)$ and the velocity field $v(x)=J(x)/\rho(x)$ are
independent of time. The continuity equation \eqref{EQcontinuity} then
implies $0=\partial\rho(x)/\partial t = -\partial[v(x)
  \rho(x)]/\partial x$, from which one obtains by spatial integration
$\rho(x)v(x)= J_0=\rm const$. Here the value of $J_0$ determines the
intensity of the flow; we recall the illustration shown in
Fig.\ \ref{FIGillustration}.

We base the machine learning procedure on a convolutional neural
network, as was done e.g.\ in Ref.~\cite{lin2019ml}, and following
Refs.~\cite{lin2019ml,lin2020ml,cats2022ml} we use many-body computer
simulations to provide training, validation, and test data. In
contrast to these equilibrium studies though, in order to address the
nonequilibrium problem we need to represent the physical time
evolution on the many-body trajectory level. We use the recently
developed highly performant adaptive BD algorithm
\cite{sammueller2021} and apply it to the three-dimensional LJ
fluid. As laid out above, in order to address situations of planar
symmetry we drive the system only along the $\ev_x$-direction. The
specific form of the driving force field $f_\ext(x)\ev_x$ is however
irrelevant, as the training data only serves to extract the intrinsic
kinematic functional relationship.

In order to cover a sufficiently broad range of flow situations, we
represent the external force field as a truncated Fourier series
$f_\ext(x)=\sum_{n=0}^{n_{\rm max}} A_n \sin(2\pi n x/L + B_n)$,
where~$L$ is the size of the cubic simulation box with periodic
boundary conditions, $A_n$ are random amplitudes with zero mean and
uniform distribution inside of a given finite interval, and $B_n$ are
random phases.  We truncate at order $n_{\rm max}=4$ such that the
length scale $L/(2\pi n_{\rm max})$ is comparable to the LJ molecular
size~$\sigma$. Ten percent of our simulation runs are carried out in
equilibrium, i.e.\ for $A_0=0$.  We use $N=500$ LJ particles inside of
a cubic simulation box of size $L=10\sigma$. The temporal duration of
each run is $1000\tau$, where $\tau=\sigma^2/D_0$ is the Brownian time
scale. After initialization the system is randomized for $1\tau$ at a
very high temperature. Then we wait for~$100\tau$ to allow the system
to reach a steady state and then collect data during the remaining
time. In total we use 1000 such simulation runs; these are subdivided
for purposes of training (520), validation (280) and testing
(200). {\ms The maximal current encountered during training was
  $J_0\sigma^2\tau=4.93$.  A~more detailed account of our
  machine-learning strategy is given in appendix \ref{SECappendix}}.

Our aim is to machine-learn and hence to explicitly demonstrate the
kinematic map, $\rho(\rv),\vel(\rv)\to \fv_{\rm int}(\rv)$ in steady
state. We present the learning algorithm with inputs $\rho(x), v(x)$
and target $f_{\rm int}(x)$. The data for these three fields are
obtained from building steady state averages via the adaptive BD over
the corresponding one-body operators. We recall the microscopic
definition of the interparticle one-body force density $\Fv_{\rm
  int}(\rv)$ via \eqr{EQFintAverage} and we refer the reader to
Appendix~A of Ref.~\cite{delasheras2019customFlow} for a description
of several methods to sample the current in BD and hence obtain the
overdamped velocity profile $\vel(\rv)$. Finally, we use the standard
counting method for the density profile~$\rho(\rv)$, although more
efficient ``force sampling'' methods \cite{rotenberg2020, borgis2013,
  delasheras2018forceSampling, renner2023torqueSampling} exist.  At
this stage we neither impose adiabatic-superadiabatic splitting
\eqref{EQFintSplitting}, nor structure-flow splitting
\eqref{EQflowStructureSplitting}, nor do we use any analytical model
form of the functional relationship. We rather work on the level of
the bare one-body simulation data, generated in the above described
randomized uniaxial flow situations of the desired planar symmetry.

We refer to the result of this procedure as the machine-learned
internal force field $f_{\rm int}^\DL(x,[\rho,v])$. This represents a
``surrogate model'' in the sense of the terminology of the machine
learning community. By construction this data structure depends
functionally on the density profile and on the velocity
profile. Importantly the external force field $f_\ext(x)$, as given by
the above described randomized Fourier series, has not been used in
the training, which was rather based solely on the intrinsic force
field and its kinematic dependence on the density profile and the
velocity field.

In order to test the validity of the functional relationship and to
address the question whether $f_{\rm int}^\DL(x,[\rho,v])$ indeed
represents the true $\fv_{\rm int}(\rv,t,[\rho,v])$ of power
functional theory, as restricted to the present planar and steady
situation, we consider a toy flow situation as an application.  We
choose the density profile to consist of a single (co)sinusoidal
deviation from the bulk, $\rho(x)=[0.5+0.2\cos(2\pi
  x/L)]\sigma^{-3}$. In order for the system to be in steady state,
the velocity then necessarily needs to satisfy $v(x)=J_0/\rho(x)$,
where the strength of the current $J_0=\rm const$ is a free parameter.

We proceed in two ways. First, we check for self-consistency.
Therefore we solve the force density balance relationship
\eqref{EQcurrentBD} for the external force field, which yields the
explicit result:
\begin{align}
  f_\ext(x) &= 
  k_BT \frac{\partial \ln\rho(x)}{\partial x}
  + \gamma v(x)
  - f_{\rm int}^\DL(x,[\rho,v]).
  \label{EQinstantCustomFlow}
\end{align}
As is explicit in \eqr{EQinstantCustomFlow}, inputting the toy state
$\rho(x),\, v(x)$ on the right hand side yields a concrete machine
learning prediction for the external force field on the left hand
side.  We then input this result for $f_\ext(x)$ as the driving force
field in a single adaptive BD simulation run and expect this procedure
to reproduce the density and velocity profile of the toy state. The
reproductive success will however materialize only provided that i)
the functional kinematic dependence actually exists and that ii) it is
accurately represented by the neural network.

The results, shown in Fig.~\ref{FIGprofiles}, demonstrate the
accomplishment of the reconstruction of the toy state. This
establishes that the machine learned functional provides a numerically
highly accurate representation of the true internal force
functional. 
{\ms That quantitative differences between results from direct BD and
  from machine learning occur for the case of strongest flow
  ($J_0\sigma^2\tau=5$) is not surprising, given that the value of the
  current is beyond the maximum encountered during training
  ($J\sigma^2\tau=4.93$). However, despite the quantitative deviations
  of the prediction for the interparticle force field, the qualitative
  behaviour of the network remains entirely reasonable.}

We take this validation via the machine learning to be a practical,
data-science-level verification of the existence of the power
functional kinematic map. We recall the original formal construction
\cite{schmidt2013pft,schmidt2022rmp} and its subsequent confirmation
via custom flow
\cite{delasheras2019customFlow,renner2021customFlowMD}.

Turning to the physics of the compressional flow, we use the
adiabatic-superadiabatic decomposition \eqref{EQFintSplitting}
together with the flow-structure splitting
\eqref{EQflowStructureSplitting} to analyze both the machine-learned
functional $f_{\rm int}^\star(x,[\rho,v])$ as well as the direct
simulation results.  As anticipated, both flow and structural force
fields have nontrivial spatial variation, see
Fig.~\ref{FIGprofiles}. The flow force primarily contains viscous
effects that stem from the dissipation that the compressional and
extensional regions of the flow pattern generate. The structural force
field becomes more strongly inhomogeneous and also larger in magnitude
upon increasing the amplitude of the flow. This trend is necessary to
provide a balance for the increasingly asymmetric and growing external
force field, which in turn is required to keep the density profile
unchanged upon increasing the throughput through the prescribed
density wave. 

The power functional predictions \eqref{EQsupForceFlow} and
\eqref{EQsupForceStructural} capture these effects reasonably well
given the simplicity of the analytical expressions, see the insets in
Fig.~\ref{FIGprofiles}. We find our numerical results to satisfy the
Noether sum rules \eqref{EQsumRuleFsup} and \eqref{EQsumRuleFflow} to
very good accuracy.  {\ms The values of the prefactors~$\eta$ and
  $\chi$ in Eqs.~\eqref{EQsupForceFlow} and
  \eqref{EQsupForceStructural} characterize the dominant behaviour of
  the system in response to spatial variation of the flow. The
  parameter $\eta/(\eps \tau\sigma^3 )=0.35$ measures viscosity and
  $\chi/(\eps \tau^2 \sigma^6 ) =0.075$ quantifies the strength of
  nonequilibrium structure formation. We regard these amplitudes as
  being well-defined transport coefficients, which will determine the
  leading behaviour of the system in situations where higher-order
  gradient contributions are small or even irrelevant.  Using our
  methodology, the precise values of $\eta$ and $\chi$ can be obtained
  systematically from straightforward comparison to the data from the
  BD simulations or the machine learning model.}

It remains to point out the stark contrast with the standard
DDFT~\eqref{EQddft}, which gives a trivial null result in the present
setup by construction: the density profile remains unchanged upon
increasing flow, and so does the adiabatic force field. So the DDFT
provides no mechanism to account for the genuine nonequilibrium
physics; {\ms see appendix~\ref{SECappendix} for further details.}

\section{Conclusions}
\label{SECconclusions}

For the purpose of assessing the status of the DDFT equation of motion
\eqref{EQddft} we have first described two exact limits that this
approximation reproduces: the dynamics of the noninteracting diffusive
ideal gas [see \eqr{EQidealDiffusionEquation}] and the spatially
inhomogeneous static equilibrium limit [see \eqr{EQdft}]. On general
grounds one expects the DDFT to perform well when the situation under
consideration is close to one of these limits. In particular near the
static case this is nontrivial, as the system might be dense and
spatially highly structured, as evident by a strongly inhomogeneous
density profile. Provided that the dynamics are driven weakly enough
via a time-dependent external potential then the DDFT
\cite{tevrugt2020review,tevrugt2022perspective} can be a highly useful
device, which enables one to describe the temporal evolution as a
chain of equilibrium states, labelled by time. {\ms As the strictly
  static case (DFT) can correctly describe arbitrary spatial
  inhomogeneities, such adiabatic time evolution can provide highly
  nontrivial information. It is challenging, however, to know {\it a
    priori} whether or not the nonequilibrium situation under
  investigation will be close to adiabatic. Leaving the use of bare
  physical intuition aside, we are not aware of any simple
  quantitative criterion that would allow one to judge a priori
  whether DDFT is reliable or not. In this sense, the DDFT
  approximation can be viewed as being uncontrolled.}

In general the contributions beyond the equilibrium physics will be
relevant. On the level of the formally exact one-body equation of
motion \eqref{EQofMotionForDensity}, the superadiabatic force field
$\fv_\sup\rt$ will then contribute and will potentially do very
significantly so. Together with the adiabatic force field, which
follows from the equilibrium excess free energy functional via
$-\nabla \delta F_\exc[\rho]/\delta\rho\rt$, their sum constitutes the
full interparticle forces. These force fields are coarse-grained, in a
microscopically sharp way, to the one-body level of dynamical
correlation functions. We have argued i) that power functional theory
is a concrete formal structure that allows to obtain $\fv_\sup\rt$
from a generating functional and ii) that simple approximate forms
already capture much relevant nonequilibrium physics and they do so in
a transparent and systematic way, and iii) that machine learning can
be used as a practical representation.

We have described and exemplified for uniaxial steady compressional
flow of the three-dimensional Lennard-Jones fluid the kinematic
functional map that governs the exact nonequilibrium dynamics on the
one-body level of dynamic correlation functions. As this description
is based on a single position coordinate and a single time variable,
it is of both conceptual and practical simplicity. As described by
power functional theory the superadiabatic interparticle force field
functionally depends on the density and the velocity field,
i.e.\ $\fv_\sup(\rv,t,[\rho,\vel])$, for overdamped Brownian
motion. The functional dependence is causal, i.e.\ on the values of
the density profile and velocity field at previous times, in general
up to an initial state. The superadiabatic force field carries this
kinematic dependence, i.e.\ on the history of $\rho\rt$ and $\vel\rt$,
but crucially it is independent of the external force field that
drives the system.

We have explicitly demonstrated the functional map
$\rho\rt,\vel\rt\to\fv_{\rm int}\rt$ by establishing this functional
relationship via machine learning the intrinsic force field.  {\ms
  This includes as a special case the equilibrium map $\rho(\rv,t)\to
  \fv_{\rm ad}(\rv,t)$, as it is relevant for the approximative
  adiabatic time evolution via the DDFT \eqref{EQddft}.  } Using the
force balance then gives direct access to the form of the required
external force field via \eqr{EQinstantCustomFlow}.  The
machine-learned model of the functional map hence enables ``instant
custom flow'' at negligible computational cost at the time of use. We
recall that the custom flow method \cite{delasheras2019customFlow,
  renner2021customFlowMD} is based on the kinematic functional map,
such that from knowing the kinematic one-body fields, the external
force field that is necessary to generate the given time evolution
follows straightforwardly from the exact force
balance~\eqref{EQcurrentBD}.
%%%

An analytical approach to one-body functional maps leads to the simple
structure of velocity gradient forms for the viscous and structural
superadiabatic forces, as exemplified in Eqs.~\eqref{EQPtexcFlow} and
\eqref{EQPtexcStructure} for compressional flow, i.e.\ for velocity
fields with nonvanishing divergence. As we have shown, the resulting
predictions for the flow force \eqref{EQsupForceFlow} and for the
structural force field \eqref{EQsupForceStructural} represent a
reasonable description of the simulation data and its representation
via the machine-learned functional. We attribute the remaining
differences to higher-order terms \cite{delasheras2020fourForces}
which we have not addressed here for simplicity {\ms and which we
  expect to become increasingly relevant under stronger driving.}  As
we have shown, our results from direct simulation, from machine
learning, and from the analytical approximations, satisfy exact global
Noether sum rules.

We have restricted our discussion to a single and relatively easily
accessible type of nonequilibrium dynamics, that of stationary
uniaxial compressional flow that represents a model steady (batch)
sedimentation situation. The power functional approach allows to go
much further, including the treatment of viscoelasticity
\cite{treffenstaedt2020shear}, as arising from superadiabatic memory,
deconfinement under shear \cite{jahreis2019shear}, the dynamic decay
of the van Hove pair correlation function as governed by drag, viscous
and structural forces
\cite{treffenstaedt2021dtpl,treffenstaedt2022dtpl}, and the complex
forms of both flow and structural forces that arise under spatially
complex forms of driving \cite{delasheras2020fourForces}.
Time-dependent uniaxial flow is relevant in a variety of situations,
including colloidal stratification \cite{he2021,kundu2022ddft} and
sedimentation \cite{sui2018}.

Although power functional theory operates on the one-body level of
dynamical correlation functions, two-body correlation functions are
accessible both formally via the nonequilibrium Ornstein-Zernike route
\cite{schmidt2022rmp} and explicitly by the dynamical test particle
limit. The latter is the dynamic generalization of Percus' static test
particle limit \cite{percus1962}, which identifies two-point
correlation functions, such as $g(r)$ as also recently shown to be
intimatedly related to thermal Noether invariance at second order
\cite{sammueller2023whatIsLiquid}, with one-body density profiles in
an external potential. This is set equal to the interparticle pair
potential.

The dynamical test-particle limit goes further in that it describes
the test particle via its own dynamical degrees of freedom, which are
coupled to those of all other particles in the system.  The concept
was originally formulated as an approximation within DDFT
\cite{archer2007dtpl,hopkins2010dtpl} and formally exactly within
power functional theory \cite{brader2015dtpl}.  Two-body
superadiabatic effects were shown via simulation work to be
significant \cite{schindler2016dynamicPairCorrelations,
  treffenstaedt2021dtpl, treffenstaedt2022dtpl} and they arise
naturally in an exact formulation of the test particle dynamics
\cite{brader2015dtpl}. The test particle limit allowed for a
rationalization of the dynamical pair structure as
e.g.\ experimentally observed in two-dimensional
colloids~\cite{stopper2018dtpl}. Recently an approach to DDFT based on
the two-body level was formulated \cite{tschopp2022ddft2} {\ms and
  earlier work was addressed at construction of dynamical density
  functional theories from exactly solvable limits \cite{scacchi2016};
  we recall that Refs.~\cite{tevrugt2020review,
    tevrugt2022perspective} provide exhaustive overviews.}
In event-driven BD simulations superadiabatic forces were shown to
consist of drag, viscous, and structural contributions
\cite{treffenstaedt2021dtpl,treffenstaedt2022dtpl}; see
Ref.~\cite{schmidt2022rmp} for an extended discussion. The physics of
active particles \cite{krinninger2016,krinninger2019,
  hermann2019pre,hermann2019prl,hermann2021molPhys} is very
significantly governed by a vigorous interplay between superadiabatic
and adiabatic forces, both of which are very strong, as the tendency
of these systems to self-compress leads naturally to very high local
densities.

Furthermore, relevant and interesting microscopic models that go
beyond the simple fluid paradigm of a pair potential, such as the
monatomic water model by Molinero and Moore
\cite{molinero2009,coe2022water} and the three-body gel by Saw et
al.\ \cite{saw2009,saw2011}, are accessible. Despite the complexity of
both its defining Hamiltonian and the intricate transient network
structure, the inhomogeneous viscous response of the three-body gel
was recently demonstrated \cite{sammueller2022gel} to be surprisingly
well captured by a simple power functional flow approximation.  We
finally recall that superadiabatic effects transcend overdamped
dynamics, and are relevant both in quantum dynamics
\cite{schmidt2015qpft, bruetting2019viscosity, schmidt2022rmp} and in
classical molecular dynamics
\cite{schmidt2018md,renner2022prl,schmidt2022rmp}.

While we have restricted ourselves to discussing the point of view of
functional relationships, it would be interesting to explore in future
work possible cross connections to other theoretical approaches, such
as Onsager's variational principle for soft matter \cite{doi2011,
  doi2015, wang2021onsager,wang2022prl}, stochastic thermodynamics
\cite{seifert2012}, large deviation theory \cite{jack2010,jack2015},
mode-coupling theory \cite{janssen2015,janssen2018}, generalized
hydrodynamics \cite{mazzuca2023}, {\ms local molecular field theory
  for nonequilibrium systems \cite{rodgers2020}}, as well as to the
physics of nonequilibrium phase transitions \cite{lips2018}, Brownian
solitons \cite{antonov2022}, {\ms and crystal dynamics
  \cite{haussmann2016,haussmann2022,ganguly2022,lin2021dcf} and
  non-isothermal situations \cite{anero2013}.  Implications of the
  machine-learning methodology are summarized at the end of
  appendix~\ref{SECappendix}. }

\acknowledgments

This work is supported by the German Research Foundation (DFG) via
Projects No.\ 436306241 and No.\ 447925252.

\appendix
{\ms \section{Simulation and training details}}
\label{SECappendix}

\begin{figure*}[htb!]
  \includegraphics[width=0.99\linewidth]{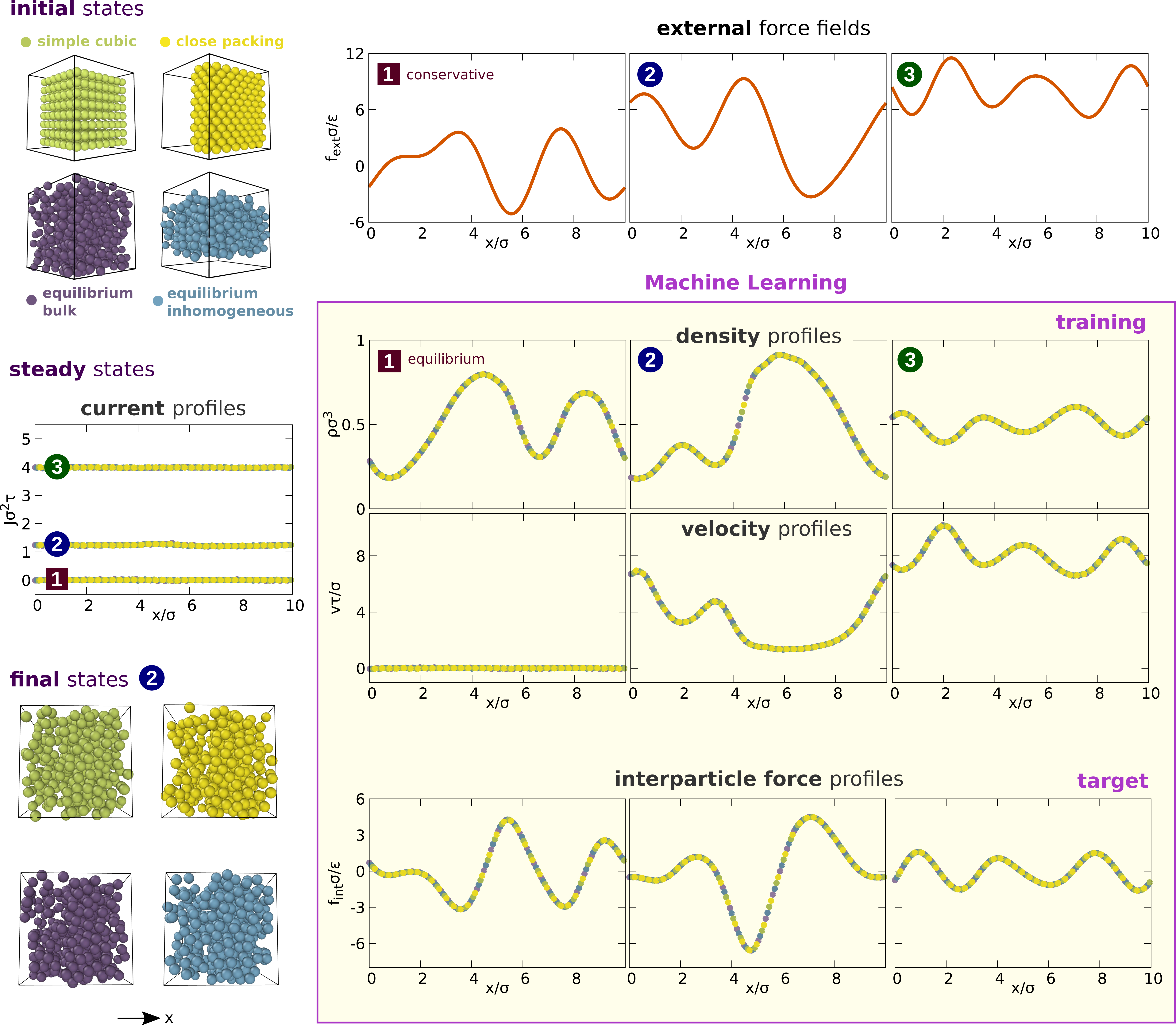}
  \caption{
    Three representative cases (systems 1, 2, and 3) of nonequilibrium
    steady states under external driving, as used for training the
    neural network. The results are obtained from adaptive~BD
    simulations under the influence of a temporally static external
    force field $f_\ext(x)$ (top row) that consists of a random
    superposition of spatial Fourier modes. The zeroth Fourier mode
    represents a constant force offset, which either vanishes (system
    1) or drives the system out of equilibrium (systems 2 and 3).
    The steady states are characterized by spatially and temporally
    constant currents $J(x)=J_0=\rm const$, with $J_0=0$ in
    equilibrium (system 1) and $J_0>0$ under drive (systems 2 and
    3). The density profile $\rho(x)$ (second row) is temporally
    constant and spatially inhomogeneous, as induced by the action of
    the external force field. The inhomogeneous velocity profile is
    simply the inverse $v(x)=J_0/\rho(x)$ (third row).
    For each system 1, 2 and 3, the results for the respective steady
    state profiles $\rho(x), v(x)$, and the interparticle force field
    $f_{\rm int}(x)$ (fourth row) are independent of the type of
    initialization, as demonstrated by the results from the four
    differently initialized simulation runs (indicated by differently
    coloured symbols) lying on top of each other.
    Here each system is alternatively initialized via one of four
    protocols to select the initial microstate: simple cubic (green)
    or close packed (yellow) crystals or an equilibrated bulk liquid
    state (violet) or an inhomogeneous slab-like confined liquid
    (blue).
    Representative simulation snapshots in steady state (``final
    states'' after 1000$\tau$) for system 2 illustrate the spatial
    structure of the flowing liquid; no imprints of the initialization
    can be perceived.
    The machine learning is based on training data $\rho(x)$ and
    $v(x)$ [or alternatively to the latter $J(x)$ or $\partial
      v(x)/\partial x$] that lead to the target interparticle force
    field $f_{\rm int}(x)$ for each given training system (1000 in
    total) in the supervised learning.
    While we here solely illustrate the training procedure, we show in
    Figs.~\ref{FIGprofiles}, \ref{FIGwaves}, and \ref{FIGddft} how the
    trained model can be used to both predict and design
    nonequilibrium steady states.
    \label{FIGtraining}
  }
\end{figure*}

\begin{figure*}[htb!]
  \includegraphics[width=0.99\linewidth]{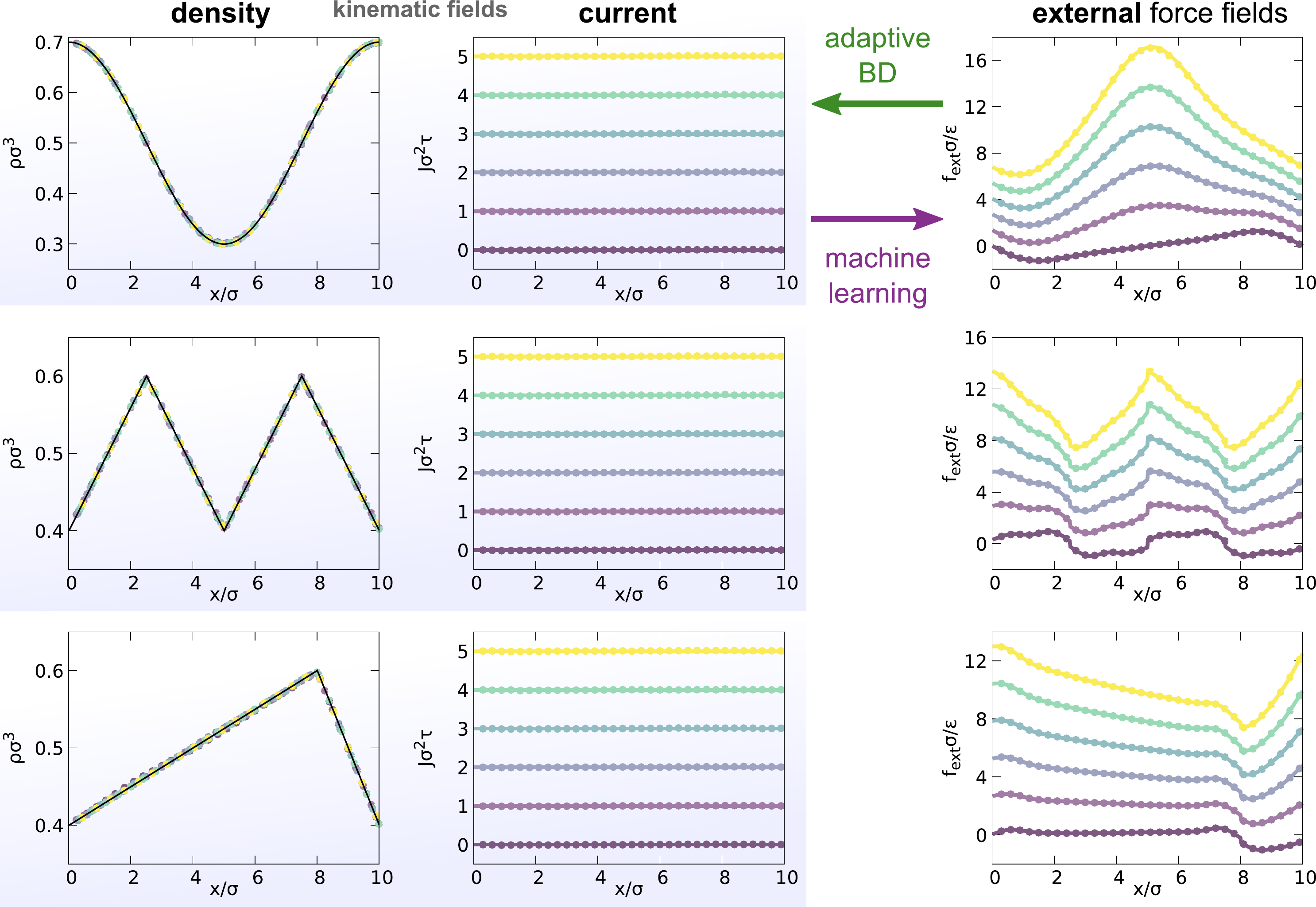}
  \caption{Demonstration of the machine-learned kinematic force map
    and its use in the design of nonequilibrium steady states. Shown
    are three different target shapes of the density profile $\rho(x)$
    (left column): cosine wave (top row, results are identical to
    those shown in Fig.~2), triangle wave (second row), and
    sawtooth-like wave (third row). For each of the three density
    waves the current profile (second column) $J(x)=J_0=\rm const$,
    and we prescribe alternative values $J_0 \tau/\sigma^2=0$
    (equilibrium), 1, 2, 3, 4, and 5 (all nonequilibrium).
     Our aim is to obtain the specific form of $f_\ext(x)$ via our
     trained machine learning model which then generates the target
     shapes of $\rho(x)$ and $J(x)$. That this procedure indeed is
     successful is validated with BD simulations as shown via the
     symbols, as both $\rho(x)$ and $J(x)$ are reproduced to high
     accuracy.
    Each of the three target density profiles $\rho(x)$ (lines) is
    strikingly matched by the simulation results (symbols);
    the data for the external force profile $f_\ext(x)$ (right column)
    conincide per construction, as the output from machine learning is
    taken as the input force field in the many-body simulations that
    are carried out to assess the accuracy of the design procedure.
    \label{FIGwaves}
  }
\end{figure*}

\begin{figure}[htb!]
    \includegraphics[width=0.99\linewidth]{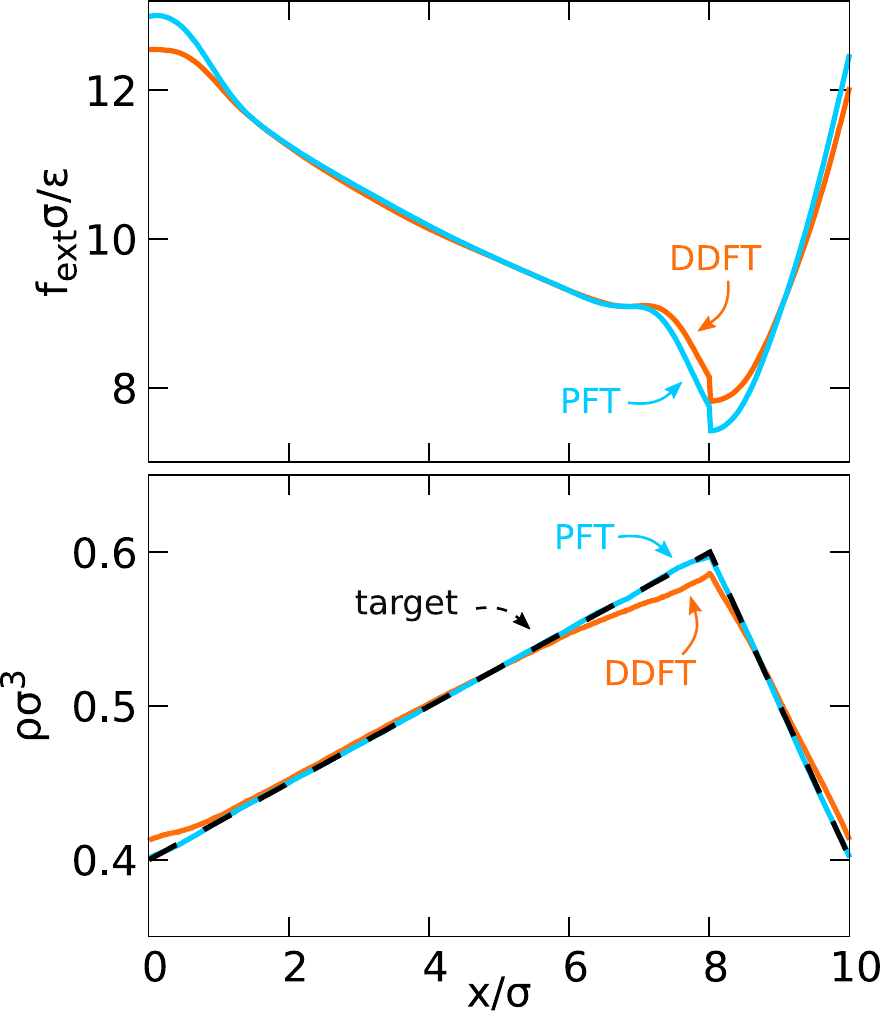}
  \caption{Shape of the external force field $f_\ext(x)$ (upper panel)
    that stabilizes the target nonequilibrium sawtooth density profile
    $\rho(x)$ (lower panel) with current strength $J_0
    \sigma^2/\tau=5$ (as also shown in the third row of
    Fig.~\ref{FIGwaves}).
    The force field is generated either from power functional
    theory via Eq.~\eqref{EQappendixInstantCustomFlow} or from machine
    learning DDFT via
    Eq.~\eqref{EQappendixInstantAdiabaticCustomFlow}.
    While capturing the correct shape of the external force profile,
    the DDFT produces in this case a quantitative error between
    maximal and minmal force of size $\epsilon/\sigma$.
    The respective results for the density profiles are obtained from
    direct simulations.  The power functional design meets the target
    profile in a quasi-exact way, whereas quantitative deviations
    occur in DDFT in particular near the cusps of the wave.
    \label{FIGddft}
  }
\end{figure}

{\bf Initialization.} The nonequilibrium many-body physics that we
investigate falls into the class of temporal initial value
problems. This holds true both on the full many-body (phase space)
level, as accessible via the simulations or formally via the
Smoluchowski equation \cite{schmidt2022rmp}, as well as on the reduced
one-body level of the temporal kinematic fields, i.e.\ the
time-dependent density profile and current distribution. Evidently
one-body initial data needs to be available in order to start the time
evolution according to either the approximate DDFT \eqref{EQddft} or
the formally exact power functional equation of motion
\eqref{EQofMotionForDensity}. In typical applications of one of these
frameworks to a dynamical problem, the system is taken to be in an
equilibrium state at a starting time~$t_0$ such that the current
vanishes and the density profile is known.
Standard ways \cite{zwanzig2001} to specify an initial state include
choosing the bulk and possibly applying (simple) external fields and
letting the system relax therein. The DFT point of view allows for
more general initializations as one can choose an equilibrium system
with an arbitrary spatially inhomogeneous density profile
$\rho(\rv,t_0)$ as the starting point.

We recall from Sec.~\ref{SECintroduction} that for an equilibrium
system with given interparticle interactions, knowledge of the density
profile [in the present case $\rho(\rv,t_0)$] is sufficient to
formally exactly determine all static thermal properties.  This fact
\cite{mermin1965, evans1979} constitues one of the major virtues of
static DFT. Leaving representability issues aside, prescribing a
(physically sensible) form of $\rho(\rv,t_0)$ is feasible, as one can
picture this as being generated by an appropriate form of
corresponding external potential $V_\ext(\rv,t_0)$, where $t_0$ acts
as a mere label to specify the initial state. As laid out above in
Sec.~\ref{SECintroduction}, the description formally requires to have
access to the free energy density functional $F[\rho]$, which
implicitly contains the full static information about the thermal
physics of the system.

In particular, the equilibrium many-body probability distribution
function $\Psi(\rv^N,t_0)$ is uniquely determined, for given
interparticle interactions and given knowledge of the density profile
\cite{evans1979, hansen2013, mermin1965}. (In the notation we have
dropped the dependence on momenta, which is trivial in the present
situation.)
One pictures the system to have been in the same equilibrium state
also at all prior times $t<t_0$ and having undergone time evolution in
this quiescent state up to $t_0$. As an aside, the equilibrium
dynamics can then be characterized for bulk liquids on the two-body
level by the van Hove dynamical correlation function
\cite{treffenstaedt2021dtpl, treffenstaedt2022dtpl, vanhove1954,
  hansen2013, brader2013noz, brader2014noz, archer2007dtpl,
  hopkins2010dtpl, stopper2015jcp, stopper2015pre, brader2015dtpl,
  schindler2016dynamicPairCorrelations}, with much recent progress
from the power functional point of view \cite{treffenstaedt2021dtpl,
  treffenstaedt2022dtpl}.

Obtaining a statistical description requires in principle to average
over the initial distribution of microstates.  In the context of
many-body simulations, in practice this necessitates to carry out a
sufficient number of independent realizations of the time evolution
that is under consideration. Representative studies relied on
e.g.\ $10^4$ realizations for the transient dynamics of hard spheres
under a temporally switched shear field \cite{treffenstaedt2020shear},
and on $2\times 10^6$ realizations \cite{renner2022prl} for
investigating superadiabatic acceleration effects that occur in
Molecular Dynamics. There are also special cases, such as test
particle concepts that allow efficient sampling of the bulk van Hove
function via building moving averages \cite{treffenstaedt2021dtpl,
  treffenstaedt2022dtpl}.  We also point out references
\cite{vanleeuwen1999, maitra2002} which address the initial state
dependence in the context of quantum mechanical time-dependent density
functional theory.\\

{\bf Steady states.} In the present model situation of uniaxial
compressional steady flow (we recall its graphical illustration in
Fig.~\ref{FIGillustration}) there is no need to temporally resolve the
one-body fields, as these are invariant in time. Hence we proceed in
the standard way of replacing the average over an ensemble of inital
states with a temporal average over a single trajectory of
sufficiently long duration. Recalling the details that are given in
Sec.~\ref{SECmachineLearning}, we average over time evolutions each
with duration $1000 \tau$, with $\tau=\sigma^2/D_0$ denoting the
Browian time scale for self diffusion with diffusion constant
$D_0=k_BT/\gamma$, where $\gamma$ is the friction constant against the
background. The strategy of identifying temporal and ensemble averages
relies on having an ergodic system of which the time evolution indeed
explores the entirety of the ensemble. As we deal with a liquid in the
present illustrative case, we expect ergodicity to hold.

In order to validate this expectation, we have investigated the
possibility of dependence of the steady state results on the initial
state of the simulation; see Fig.~\ref{FIGtraining} for illustrations
of the chosen four different (crystalline and disordered) microstates.
We recall that we evolve the system over a waiting time of $100\tau$
before starting to sample the one-body correlation functions. The
resulting steady state profiles, see Fig.~\ref{FIGtraining}, bear no
traces of the different initialization, and the data of each of the
four runs collapse onto each other. As this behaviour is already
observed on the level of starting with individual differing
microstates, we expect no changes if we were to start with a
representative sample of, say $10^4$, microstates in order to
numerically approximate an entire distribution.\\

{\bf Training procedure.} The results shown in Fig.~\ref{FIGtraining}
also serve to illustrate our training procedure in more detail. We use
randomized forms of the external force field $f_\ext(x)$ via
superimposing Fourier modes that are compatible with the box size $L$
via: $f_\ext(x)=\sum_{n=0}^{n_{\rm max}} A_n \cos(2\pi n x/L + B_n)$,
which we cut off at $n_{\rm max}=4$. The amplitudes $A_n$ and phases
$B_n$ are generated randomly within a cutoff, which i) makes the
specific training protocol free of having to perform manual choices
and ii) removes any further bias thus easing the interpretation of the
quality of the machine learning predictions.
Our currently adopted random training strategy is suited to
investigate issues of generality, universality and transferability
both of the underlying mathematical structure of power functional
theory and of its presently proposed specifically tailored
implementation via supervised machine learning.
Nevertheless, as our supervised learning
procedure is general, one could well tune for specific applications
and rather train on the basis of situations that are close to the
eventual use of the network, see e.g.\ Ref.~\cite{cats2022ml} for a
corresponding enlightening study.

Our supervision protocol operates on the level of the simulation
output in a specifically organized way of (functional) dependencies of
the obtained histograms that represent the one-body distributions. All
three involved functions are taken from straightforward sampling in
adaptive BD. The density profile $\rho(x)$ is obtained from the
standard counting method. (We recall reduced-variance sampling
techniques such as force sampling
\cite{delasheras2018forceSampling,rotenberg2020, borgis2013}, which
could help in acquiring numerical training data more efficiently.)

The current distribution $J(x)$ can be sampled via the force balance
equation or alternatively via a temporally centered derivative of the
particle trajectories. Here the position resolved histogram is filled
with the displacement vector of each particle, between the position in
the next and in the previous time step; see the appendix of
Ref.~\cite{delasheras2019customFlow} for an in-depth description. The
velocity profile then follows straightforwardly from
$v(x)=J(x)/\rho(x)$. 

Finally the interparticle force density $F_{\rm int}(x)$ is sampled in
an analogous way on the basis of a position-resolved histogram that
simply accepts the instantaneous interparticle force that acts on each
given particle. The interparticle force field is then obtained by
simple normalization with the density profile, $f_{\rm int}(x)=F_{\rm
  int}(x)/\rho(x)$.

From a data science point of view, and possibly even when working in a
physics-informed way, one might use the information in all three
fields $\rho(x)$, $v(x)$ and $f_{\rm int}(x)$ to analyze and make
predictions of the dynamical behaviour of the system. However, our
present approach is very specific and leaves no choice in the general
setup of the supervised learning. We have $\rho(x)$ and $v(x)$ as
inputs and $f_{\rm int}(x)$ as the output or target of the neural
network; we recall the description in Sec.~\ref{SECpft} of the
functional relationships that apply to the nonequilibrium physics. As
concrete illustration we show three representative training data sets
in Fig.~\ref{FIGtraining}. We find the success of the learning
procedure to be robust against changes in even simple network topology
and choice of hyperparameters.  We attribute these features to the
fact that the mapping is formally exact, as summarized in the main
text and reviewed in detail in Ref.~\cite{schmidt2022rmp}.\\

{\bf Universality and inverse design.} 
The thus obtained functional relationship $f_{\rm
  int}^\star(x,[\rho,v])$ not merely interpolates between the training
situations, but it rather captures the genuine correlated nature of
the statistical physics under consideration. We recall the target
cosine-shaped density wave (presented in Fig.~\ref{FIGprofiles} and
replotted in the first row of Fig.~\ref{FIGwaves} as a
reference). This situation was not genuinely part of the training but
certainly close in character. As a further demonstration we apply the
network to a triangular density wave (second row in
Fig.~\ref{FIGwaves}) and also to a sawtooth-like density wave (third
row in Fig.~\ref{FIGwaves}), which are further removed from the
situations encountered during training. The excellent match with the
direct simulation results of the density profile and current
distribution validates our approach.

We finally return to the DDFT and to the purpose of the present
Perspective of discussing its virtues and shortcomings. We present as
a final example, see Fig.~\ref{FIGddft}, its application to the
sawtooth state. Shown is the external force profile, as obtained from
the machine-learned power functional, via the instant custom flow
Eq.~\eqref{EQinstantCustomFlow}, which we reproduce for convenience:
\begin{align}
  f_\ext(x) & = \gamma \frac{J_0}{\rho(x)} 
  + k_BT\frac{\partial \ln \rho(x)}{\partial x}
  - f_{\rm int}^\star(x,[\rho,v]).
  \label{EQappendixInstantCustomFlow}
\end{align}
In the present case this constitutes a (numerical) quasi-exact
solution; we recall the excellent agreement of the resulting kinematic
profiles, shown in Fig.~\ref{FIGwaves}, with the direct simulation
results. We can now easily compare this to the adiabatic DDFT
prediction, as the equilibrium states were part of our training
protocol (cases of vanishing zeroth mode of the external force
field). We can straightforwardly implement this on the level of the
neural network by simply considering no flow, i.e.\ setting the
velocity profile $v(x)=0$, which yields a quasi-exact representation
of the adiabatic force; we recall the perfect agreement shown in
Fig.~\ref{FIGwaves} for the cases with no flow, where $J_0=0$ (dark
purple symbols and lines). Hence within DDFT the instant custom flow
Eq.~\eqref{EQappendixInstantCustomFlow} reduces to the following
simple form:
\begin{align}
  f_\ext^{\rm DDFT}(x) = \gamma \frac{J_0}{\rho(x)}
  +k_BT \frac{\partial \ln \rho(x)}{\partial x}
  -f_{\rm ad}^\star(x,[\rho]).
%  -f_{\rm int}^\star(x,[\rho,v=0]).
  \label{EQappendixInstantAdiabaticCustomFlow}
\end{align}
Here the machine-learned adiabatic force field is obtained simply by
evaluating the full nework at vanishing velocity, i.e., $f_{\rm
  ad}^\star(x,[\rho])=f_{\rm int}^\star(x,[\rho,v=0])$.  The
comparison shown in Fig.~\ref{FIGddft} indicates qualitatively correct
behaviour of the DDFT but quantitative errors { of magnitude
  $\epsilon/\sigma$.}

The effects being quantitatively small should not lead one to conclude
that the physics are irrelevant. First, as laid out above, the planar
uniaxial geometry is intrinsically favorable for the DDFT. Secondly,
we have chosen moderate values for temperature and for density to ease
our current pilot study for the use of machine learning. This renders
the situation relatively simply. However, as emphasized above, a
priori it is very difficult to assess whether or not the DDFT will be
sufficient to obtain a reliable estimate of the real time
evolution. For a very recent demonstration of quantitatively large
superadiabatic (viscous and strutural) forces, we point the reader to
the study by Samm\"uller et al \cite{sammueller2022gel} of the
nonequilibrium dynamics of a three-body colloidal gel former
\cite{saw2009,saw2011}.  \\

{\bf Implications and related work.} We take the quantitative success
and computational ease of applying supervised machine learning to the
formal functional dependencies of nonequilibrium many-body flow, as
presently considered in a simple uniaxial flow geometry, as an
incentive to summarize several possible connections that could be
explored in future work. Our approach fits into the broader picture of
coarse-graining many-body systems out of equilibrium
\cite{schilling2022} and it is very specific in terms of input and
output variables of the neural network. The supervised machine
learning method gives the interparticle force field directly, in
contrast to approaches that are based on learning the excess free
energy functional \cite{lin2019ml, lin2020ml,cats2022ml,
  yatsyshin2022} which then upon differentiation according to
Eq.~\eqref{EQFad} give the (adiabatic approximation for the)
interparticle force density. DDFT was also used in learning the
physics of pattern formation from images \cite{zhao2020}, and for
importance sampling in adaptive multiscale simulations
\cite{bhatio2021}; Ref.~\cite{tevrugt2022perspective} lists several
further examples.

In the terminology of multiscale simulation methods for soft matter
systems \cite{schmid2022editorial}, in our method we learn the
characteristics of a fine-grained model (chosen as the LJ fluid in the
present model study) and, while not strictly obtaining a
coarse-grained model, are able to reduce the fine-grained information
systematically to the one-body level in a microscopically resolved
way. It would be interesting to see whether our approach is useful in
the context of adaptive simulation techniques \cite{ebrahimi2021,
  dellesitte2019, baptista2021} as applied e.g.\ to coupling
boundaries of open systems \cite{gholami2021}.

We would need to go beyond the presently considered steady states to
address memory effects, as are relevant both in classical
\cite{jung2017reconstructing, klippenstein2021, treffenstaedt2021dtpl,
  treffenstaedt2022dtpl} and in quantum systems \cite{maitra2002,
  vanleeuwen1999}.
Given the recent interest in force-based quantum-mechanical density
functional theory \cite{tchenkoue2019, tarantino2021}, we could
imagine that apart from quantum power functional theory
\cite{hermann2022quantum, bruetting2019viscosity, schmidt2015qpft,
  schmidt2022rmp} our study could potentially be inspirational in
other force-based approaches to quantum dynamics \cite{tokatly2005one,
  tokatly2005two, ullrich2006, tokatly2007}.


\vspace{40mm}

\begin{thebibliography}{31}


\bibitem{nagel2017}
  S. R. Nagel,
  {Experimental soft-matter science},
  \href{https://doi.org/10.1103/RevModPhys.89.025002}
       {Rev. Mod. Phys. {\bf 89}, 025002 (2017).}

\bibitem{evans2019physicsToday}
  R. Evans, D. Frenkel, and M. Dijkstra, 
  {From simple liquids to colloids and soft matter},
  \href{https://doi.org/10.1063/PT.3.4135}
     {Phys. Today {\bf 72}, 38 (2019).}

\bibitem{hansen2013} 
  J.~P. Hansen and I.~R. McDonald, {\it Theory of
  Simple Liquids}, 4th ed.\  (Academic Press, London, 2013). 

\bibitem{evans1979} R. Evans, 
  {The nature of the liquid-vapour
  interface and other topics in the statistical mechanics of
  non-uniform, classical fluids},
  \href{https://doi.org/10.1080/00018737900101365}
       {Adv. Phys. {\bf 28}, 143 (1979).}

\bibitem{marconi1999}
  U. M. B. Marconi and P. Tarazona, 
  {Dynamic density functional theory of fluids},
  \href{https://doi.org/10.1063/1.478705}
  {J. Chem. Phys. {\bf 110}, 8032 (1999).}

\bibitem{hohenberg1977}
  P. C. Hohenberg and B. I. Halperin,
  {Theory of dynamic critical phenomena,}
  \href{https://doi.org/10.1103/RevModPhys.49.435}
       {Rev. Mod. Phys. {\bf 49}, 435 (1977).}

\bibitem{sammueller2021}
  F. Samm\"uller and M. Schmidt, 
  {Adaptive Brownian dynamics},
  \href{https://doi.org/10.1063/5.0062396}
       {J. Chem. Phys. {\bf 155}, 134107 (2021). }

\bibitem{royall2007dynamicSedimentation}
   C. P. Royall, J. Dzubiella, M. Schmidt, and A.~van Blaaderen, 
   {Non-equilibrium sedimentation of colloids on the particle scale,}
   \href{https://doi.org/10.1103/PhysRevLett.98.188304}
        {Phys. Rev. Lett. {\bf 98}, 188304 (2007).}

\bibitem{bier2008prl}
    M. Bier, R. van Roij, M. Dijkstra, and P. van der Schoot,
    {Self-diffusion of particles in complex fluids:
      temporary cages and permanent barriers,}
    \href{https://doi.org/10.1103/PhysRevLett.101.215901}
         {Phys. Rev. Lett. {\bf 101}, 215901 (2008)}.

\bibitem{stopper2018dtpl}
   D. Stopper, A. L. Thorneywork, R. P. A. Dullens, and R. Roth,
   {Bulk dynamics of Brownian hard disks: Dynamical density functional
     theory versus experiments on two-dimensional colloidal hard spheres,}
   \href{https://doi.org/10.1063/1.5019447}
        {J. Chem. Phys. {\bf 148}, 104501 (2018).}


{\ms \bibitem{perez2021}
  C. M. Perez, M. Rey, B. D. Goddard, J. H. J. Thijssen,
  Changing the flow profile and resulting drying pattern of
  dispersion droplets via contact angle modification,
\href{https://doi.org/10.48550/arXiv.2111.00464}
     {arXiv:2111.00464}.
}

\bibitem{evans1992}
  R. Evans,
  {Density functionals in the theory of nonuniform fluids},
  in {\it Fundamentals of Inhomogeneous Fluids},
  edited by D. Henderson (Dekker, New York, 1992).

\bibitem{evans2016}
  For an overview of new developments in classical density functional
  theory, see: R. Evans, M. Oettel, R. Roth, and G. Kahl, 
  {\it New developments in classical density functional theory},
  \href{https://doi.org/10.1088/0953-8984/28/24/240401}
       {J. Phys.: Condens. Matter {\bf 28}, 240401 (2016).}


\bibitem{evans2019pnas}
  R. Evans, M. C. Stewart, and N. B. Wilding,  
  {A unified description of hydrophilic and superhydrophobic surfaces in 
  terms of the wetting and drying transitions of liquids},
  \href{https://doi.org/10.1073/pnas.1913587116}
       {Proc. Nat. Acad. Sci. {\bf 116}, 23901 (2019).}


\bibitem{coe2022prl}
  M. K.  Coe, R. Evans, and N. B. Wilding,
  {Density depletion and enhanced fluctuations in water near 
    hydrophobic solutes: identifying the underlying physics,}
  \href{https://doi.org/10.1103/PhysRevLett.128.045501}
       {Phys. Rev. Lett. {\bf 128}, 045501 (2022).}

\bibitem{coe2023}
  M. K.  Coe, R. Evans, and N. B. Wilding,
  {Understanding the physics of hydrophobic solvation,}
  \href{https://doi.org/10.1063/5.0134060}
       {J. Chem. Phys. {\bf 158}, 034508 (2023).}

\bibitem{martinjimenez2017natCom} 
  D. Martin-Jimenez, E. Chac\'on, P. Tarazona, and R. Garcia, 
  {Atomically resolved three-dimensional structures of
  electrolyte aqueous solutions near a solid surface},
  \href{https://doi.org/10.1038/ncomms12164}
       {Nat. Commun. {\bf 7}, 12164 (2016).}

\bibitem{hernandez-munoz2019}
  J. Hern\'andez-Mu\~noz, E. Chac\'on, and P. Tarazona, 
  {Density functional analysis of atomic force microscopy in a 
    dense fluid},
  \href{https://doi.org/10.1063/1.5110366}
       {J. Chem. Phys. {\bf 151}, 034701 (2019).}

\bibitem{cats2021decayLength}
  P. Cats, R. Evans, A. H\"artel, and R. van Roij,
  {Primitive model electrolytes in the near and far field: Decay 
    lengths from DFT and simulations,}
  \href{https://doi.org/10.1063/5.0039619}
       {J. Chem. Phys. {\bf 154}, 124504 (2021).}

\bibitem{rosenfeld1989}
  Y. Rosenfeld,
  {Free-energy model for the inhomogeneous hard-sphere fluid mixture and 
    density-functional theory of freezing}, 
  \href{https://doi.org/10.1103/PhysRevLett.63.980}
       {Phys. Rev. Lett. {\bf 63}, 980 (1989).}.

\bibitem{roth2010}
  R. Roth,
  {Fundamental measure theory for hard-sphere mixtures: a review}, 
  \href{https://doi.org/10.1088/0953-8984/22/6/063102}
       {J. Phys.: Condens. Matter \textbf{22}, 063102 (2010)}.

\bibitem{roth2002WhiteBear}
  R. Roth, R. Evans, A. Lang, and G. Kahl, 
  {Fundamental measure theory for hard-sphere mixtures
    revisited: the White Bear version,}
  \href{https://doi.org/10.1088/0953-8984/14/46/313}
       {J. Phys.: Condens. Matter {\bf 14}, 12063 (2002).}

\bibitem{roth2006WhiteBear}
  H. Hansen-Goos and R. Roth,
  { Density functional the- ory for hard-sphere mixtures: 
    the White Bear version mark II,}
  \href{https://doi.org/10.1088/0953-8984/18/37/002}
       {J.Phys.: Condens. Matter {\bf 18}, 8413 (2006).}



\bibitem{archer2004}
  A. J. Archer and R. Evans,
  {Dynamical density functional theory and its application 
    to spinodal decomposition},
  \href{https://doi.org/10.1063/1.1778374}
  {J. Chem. Phys. {\bf 121}, 4246 (2004).}

\bibitem{chan2004}
  G. K.-L. Chan and R. Finken,
  {Time-dependent density functional theory of classical fluids,}
  \href{https://doi.org/10.1103/PhysRevLett.94.183001}
       {Phys. Rev. Lett. {\bf 94}, 183001 (2005).}

\bibitem{espanol2009}
  P. Espa\~nol and H. L\"owen,
  {Derivation of dynamical density functional theory using
    the projection operator technique,}
  \href{https://doi.org/10.1063/1.3266943}
       {J. Chem. Phys. {\bf 131}, 244101 (2009)}.

\bibitem{marconi2007}
  U. M. B. Marconi and S. Melchionna,
  {Phase-space approach to dynamical density functional theory},
  \href{https://doi.org/10.1063/1.2724823}
       {J. Chem. Phys. {\bf 126}, 184109 (2007).}

\bibitem{dzubiella2003mfddft}
  J. Dzubiella and C. N. Likos,
  {Mean-field dynamical density functional theory,}
  \href{https://doi.org/10.1088/0953-8984/15/6/102}
       {J. Phys.: Condens. Matter {\bf 15}, L147 (2003).}

\bibitem{lutsko2021reconsidering}
  J. F. Lutsko and M. Oettel,
  {Reconsidering power functional theory,}
  \href{https://doi.org/10.1063/5.0055288}
       {J. Chem. Phys. {\bf 155}, 094901 (2021).}

\bibitem{szamel2022}
  G. Szamel,
  {An alternative, dynamic density functional-like theory for 
    time-dependent density fluctuations in glass-forming fluids,}
  \href{https://doi.org/10.1063/5.0091385}
       {J. Chem. Phys. {\bf 156}, 191102 (2022).}

\bibitem{goddard2021wellposedness}
  B. D. Goddard, R. D. Mills-Williams, M. Ottobre, and G. Pavliotis,
  {Well-posedness and equilibrium behaviour of overdamped dynamic 
    density functional theory,}
  \href{https://doi.org/10.48550/arXiv.2002.11663}
       {arXiv:2002.11663}.

\bibitem{archer2006inertia}
   A. J. Archer, 
   {Dynamical density functional theory for dense atomic liquids,}
   \href{https://doi.org/10.1088/0953-8984/18/24/004}
        {J. Phys.: Condens. Matter {\bf 18}, 5617 (2006).}

\bibitem{archer2009inertia}
   A. J. Archer,
   {Dynamical density functional theory for molecular and colloidal 
     fluids: A microscopic approach to fluid mechanics,}
   \href{https://doi.org/10.1063/1.3054633}
        {J. Chem. Phys. {\bf 130}, 014509 (2009).}

\bibitem{stierle2021}
  R. Stierle and J. Gross,
  {Hydrodynamic density functional theory for mixtures
    from a variational principle and its application to droplet coalescence,}
  \href{https://doi.org/10.1063/5.0060088}
       {J. Chem. Phys. {\bf 155}, 134101 (2021).}

\bibitem{goddard2012prl}
  B. D. Goddard, A. Nold, N. Savva, G. A. Pavliotis, and S. Kalliadasis,
  {General dynamical density functional theory for classical fluids,}
  \href{https://doi.org/10.1103/PhysRevLett.109.120603}
       {Phys. Rev. Lett. {\bf 109}, 120603 (2012).}


\bibitem{rex2009epje}
  M. Rex and H. L\"owen,
  {Dynamical density functional theory for colloidal
    dispersions including hydrodynamic interactions,}
  \href{https://doi.org/10.1140/epje/i2008-10363-x}
       {Eur. Phys. J. E {\bf 28}, 139 (2009)}.


\bibitem{monchojorda2020}
  J. Dzubiella, and A. Moncho-Jord\'a,
  {Controlling the microstructure and phase behavior of confined soft
    colloids by active interaction switching,}
  \href{https://doi.org/10.1103/PhysRevLett.125.078001}
       {Phys. Rev. Lett. {\bf 125}, 078001 (2020).}

\bibitem{bley2021}
  M. Bley, J. Dzubiella, and A. Moncho-Jord\'a,
  {Active binary switching of soft colloids: stability
    and structural properties,}
  \href{https://doi.org/10.1039/d1sm00670c}
       {Soft Matter {\bf 17}, 7682 (2021).}



\bibitem{goddard2021opinion}
  B. D. Goddard, B. Gooding, G. A. Pavliotis, and H. Short,
  {Noisy bounded confidence models for opinion dynamics: the effect
    of boundary conditions on phase transitions,}
  \href{https://doi.org/10.1093/imamat/hxab044}
       {IMA J. Appl. Math. {\bf 87}, 80 (2022).}


\bibitem{tevrugt2020natComm}
  M. te Vrugt, J. Bickmann, and R. Wittkowski,
  {Effects of social distancing and isolation on epidemic spreading
    modeled via dynamical density functional theory,}
  \href{https://doi.org/10.1038/s41467-020-19024-0}
       {Nat. Commun. {\bf 11}, 5576 (2020).}

\bibitem{tevrugt2020review}
  M. te Vrugt, H. L\"owen, and R. Wittkowski, 
  {Classical dynamical density functional theory: 
    from fundamentals to applications},
  \href{https://doi.org/10.1080/00018732.2020.1854965}
       {Adv. Phys. {\bf 69}, 121 (2020).}

\bibitem{tevrugt2022perspective}
  M. te Vrugt and R. Wittkowski, 
  {Perspective: New directions in dynamical density functional theory},
  \href{https://doi.org/10.1088/1361-648X/ac8633}
       {J. Phys.: Condens. Matter  {\bf 35}, 041501 (2023).}

\bibitem{schmidt2022rmp}
  M. Schmidt, 
  {Power functional theory for many-body dynamics},
  \href{https://doi.org/10.1103/RevModPhys.94.015007}
       {Rev. Mod. Phys. {\bf 94}, 015007 (2022).}

\bibitem{schilling2022}
  T. Schilling,
  {Coarse-grained modelling out of equilibrium,}
  \href{https://doi.org/10.1016/j.physrep.2022.04.006}
  {Phys. Rep.  {\bf 972}, 1 (2022).}



\bibitem{hermann2021noether}
  S. Hermann and M. Schmidt, 
  {Noether's theorem in statistical mechanics},
  \href{https://doi.org/10.1038/s42005-021-00669-2}
       {Commun. Phys.  {\bf  4}, 176 (2021).}

\bibitem{hermann2022topicalReview}
  S. Hermann and M. Schmidt, 
  {Why Noether's theorem applies to statistical mechanics,}
   \href{https://doi.org/10.1088/1361-648X/ac5b47}
        {J. Phys.: Condens. Matter 
          {\bf 34}, 213001 (2022) (Topical Review).}

\bibitem{tschopp2022forceDFT}
  S. M. Tschopp, F. Samm\"uller, S. Hermann, M. Schmidt, and J.~M. Brader,
  {Force density functional theory in- and out-of-equilibrium},
  \href{https://doi.org/10.1103/PhysRevE.106.014115}
  {Phys. Rev. E {\bf 106}, 014115 (2022).}
  
\bibitem{sammueller2022forceDFT}
  F. Samm\"uller, S. Hermann, and M. Schmidt,
  {Comparative study of force-based classical density functional theory},
  \href{https://doi.org/10.48550/arXiv.2212.01780}
       {arxiv:2212.01780}.
  
\bibitem{hermann2022quantum}
  S. Hermann and M. Schmidt,
  {Force balance in thermal quantum many-body systems from Noether's theorem},
  \href{https://doi.org/10.1088/1751-8121/aca12d}
  {J. Phys. A: Math. Theor. {\bf 55}, 464003 (2022).
    \it (Claritons and the Asymptotics of ideas: the Physics of Michael Berry)}.

\bibitem{hermann2022variance}
  S. Hermann and M. Schmidt, 
  {Variance of fluctuations from Noether invariance},
  \href{https://doi.org/10.1038/s42005-022-01046-3}
       {Commun. Phys. {\bf 5}, 276 (2022).}
       
\bibitem{sammueller2023whatIsLiquid}
  F. Samm\"uller, S. Hermann, D. de las Heras, and M. Schmidt,
  {What is liquid, from Noether’s perspective?}
  {\ms \href{https://doi.org/10.48550/arXiv.2301.11221}
       {arxiv:2301.11221}.}



\bibitem{delasheras2019customFlow}
  D.  de las Heras, J. Renner, and M. Schmidt,
  {Custom flow in overdamped Brownian dynamics,}
  \href{https://doi.org/10.1103/PhysRevE.99.023306}
       {Phys. Rev. E {\bf 99}, 023306 (2019).}

\bibitem{renner2021customFlowMD}
  J. Renner, M. Schmidt, and D. de las Heras,
  {Custom flow in molecular dynamics,}
  \href{https://doi.org/10.1103/PhysRevResearch.3.013281}
       {Phys. Rev. Res. {\bf 3}, 013281 (2021).}


\bibitem{schmidt2013pft}
  M. Schmidt and J. M. Brader, 
  {Power functional theory for Brownian dynamics},
  \href{https://doi.org/10.1063/1.4807586}
       {J. Chem. Phys. {\bf 138}, 214101 (2013).}


\bibitem{fortini2014prl}
  A. Fortini, D. de las Heras, J.~M. Brader, and M.~Schmidt, 
  {Superadiabatic forces in Brownian many-body dynamics,}
  \href{https://doi.org/10.1103/PhysRevLett.113.167801}
       {Phys. Rev. Lett. {\bf 113}, 167801 (2014).}

\bibitem{stuhlmueller2018prl}
  N. C. X. Stuhlm\"uller, T. Eckert, D. de las Heras, and M. Schmidt, 
  {Structural nonequilibrium forces in driven colloidal systems,}
  \href{https://doi.org/10.1103/PhysRevLett.121.098002}
       {Phys. Rev. Lett. {\bf 121}, 098002 (2018).}

\bibitem{treffenstaedt2020shear}
  L. L. Treffenst\"adt and M. Schmidt,
  {Memory-induced motion reversal in Brownian liquids,}
  \href{https://doi.org/10.1039/C9SM02005E}
       {Soft Matter {\bf 16}, 1518 (2020).}

\bibitem{jahreis2019shear}
  N. Jahreis and M. Schmidt,
  {Shear-induced deconfinement of hard disks,}
  \href{https://doi.org/10.1007/s00396-020-04644-1}
       {Col. Pol. Sci. {\bf 298}, 895 (2020).}

\bibitem{delasheras2018velocityGradient}
  D. de las Heras and M. Schmidt, 
  {Velocity gradient power functional for Brownian dynamics},
  \href{https://doi.org/10.1103/PhysRevLett.120.028001}
       {Phys. Rev. Lett. {\bf 120}, 028001 (2018).}

\bibitem{delasheras2020fourForces} 
  D. de las Heras and M. Schmidt,
  {Flow and structure in nonequilibrium Brownian many-body systems},
  \href{https://doi.org/10.1103/PhysRevLett.125.018001}
       {Phys. Rev. Lett. {\bf 125}, 018001 (2020).}

\bibitem{sammueller2022gel}
  F. Samm\"uller, D. de las Heras, and M. Schmidt,
  {Inhomogeneous steady shear dynamics of a three-body colloidal gel former,}
  \href{https://doi.org/10.1063/5.0130655}
       {J. Chem. Phys.  {\bf 158}, 054908 (2023)}.
       (\href{https://publishing.aip.org/publications/journals/special-topics/jcp/colloidal-gels/}{Special Topic}
       \href{https://aip.scitation.org/topic/special-collections/gels2022?SeriesKey=jcp}{on Colloidal Gels})


\bibitem{percus1962}
  J. K. Percus,
  {Approximation methods in classical statistical mechanics,}
  \href{https://dx.doi.org/10.1103/PhysRevLett.8.462}
       {Phys. Rev. Lett. {\bf 8}, 462 (1962)}.

\bibitem{archer2007dtpl}
  A. J. Archer, P. Hopkins, and M.~Schmidt,
  {Dynamics in inhomogeneous liquids and glasses via the test particle limit,}
  \href{https://doi.org/10.1103/PhysRevE.75.040501}
       {Phys. Rev. E {\bf 75}, 040501(R) (2007).}

\bibitem{hopkins2010dtpl}
  P. Hopkins,  A. Fortini, A.~J. Archer, and M.~Schmidt, 
 {The van Hove distribution function for Brownian hard spheres:
   Dynamical test particle theory and computer simulations for bulk dynamics,}
 \href{https://doi.org/10.1063/1.3511719}
      {J. Chem. Phys. {\bf 133}, 224505 (2010).}

\bibitem{stopper2015jcp}
  D. Stopper, R. Roth, and H. Hansen-Goos,
  {Communication: Dynamical density
    functional theory for dense suspensions of colloidal hard spheres,}
  \href{https://doi.org/10.1063/1.4935967}
       {J. Chem. Phys. {\bf 143}, 181105 (2015).}

\bibitem{stopper2015pre}
  D. Stopper, K. Marolt, R. Roth, and H. Hansen-Goos,
  {Modeling diffusion in colloidal suspensions by dynamical
    density functional theory using fundamental measure theory of hard spheres,}
  \href{http://dx.doi.org/10.1103/PhysRevE.92.022151}
       {Phys. Rev. E {\bf 92}, 022151 (2015).}

\bibitem{brader2015dtpl}
  J. M. Brader and M. Schmidt,
  {Power functional theory for the dynamic test particle limit,}
  \href{https://doi.org/10.1088/0953-8984/27/19/194106}
       {J. Phys.: Condens. Matter {\bf 27}, 194106 (2015).}

\bibitem{schindler2016dynamicPairCorrelations}
  T. Schindler and M. Schmidt, 
  {Dynamic pair correlations and superadiabatic forces in a dense Brownian liquid,}
  \href{https://doi.org/10.1063/1.4960031}
       {J. Chem. Phys. {\bf 145}, 064506 (2016).}

\bibitem{treffenstaedt2021dtpl}
  L. L. Treffenst\"adt and M. Schmidt,
  {Universality in driven and equilibrium hard sphere liquid dynamics,}
    \href{https://doi.org/10.1103/PhysRevLett.126.058002}
         {Phys. Rev. Lett. {\bf 126}, 058002 (2021).}

\bibitem{treffenstaedt2022dtpl}
  L. L. Treffenst\"adt, T. Schindler, M. Schmidt,
  {Dynamic decay and superadiabatic forces in the van Hove dynamics of 
    bulk hard sphere fluids},
  \href{https://doi.org/10.21468/SciPostPhys.12.4.133}
       {SciPost Phys. {\bf 12}, 133 (2022).}

\bibitem{hermann2019prl}
  S. Hermann, D. de las Heras, and M. Schmidt, 
  {Non-negative interfacial tension in phase-separated
    active Brownian particles,}
  \href{https://doi.org/10.1103/PhysRevLett.123.268002}
       {Phys. Rev. Lett. {\bf 123}, 268002 (2019).}

\bibitem{hermann2019pre}
  S. Hermann, P. Krinninger, D. de las Heras, and M. Schmidt,
  {Phase coexistence of active Brownian particles,}
  \href{https://link.aps.org/doi/10.1103/PhysRevE.100.052604}
       {Phys. Rev. E {\bf 100}, 052604 (2019).}

\bibitem{krinninger2016}
  P. Krinninger, M. Schmidt, and J. M. Brader, 
  {Nonequilibrium phase behaviour from minimization of free 
    power dissipation},
  \href{http://dx.doi.org/10.1103/PhysRevLett.117.208003}
       {Phys. Rev. Lett. {\bf 117}, 208003 (2016).}

\bibitem{krinninger2019}
  P. Krinninger and M. Schmidt, 
  {Power functional theory for active Brownian particles: 
    general formulation and power sum rules,}
  \href{https://doi.org/10.1063/1.5061764}
       {J. Chem. Phys. {\bf 150}, 074112 (2019).}

\bibitem{hermann2021molPhys}
  S. Hermann, D. de las Heras, and M. Schmidt, 
  {Phase separation of active Brownian particles in two dimensions: 
    Anything for a quiet life,}
  \href{https://doi.org/10.1080/00268976.2021.1902585}
       {Mol. Phys. e1902585 (2021).}


\bibitem{delasheras2014canonical}
  D. de las Heras, M. Schmidt,
  {Full canonical information from grand potential density 
    functional theory,}
  \href{https://doi.org/10.1103/PhysRevLett.113.238304}
       {Phys. Rev. Lett. {\bf 113}, 238304 (2014).}

\bibitem{delasheras2016particleConserving}
  D. de las Heras, J. M. Brader, A. Fortini, M. Schmidt,
  Particle conservation in dynamical density functional theory,
  \href{https://doi.org/10.1088/0953-8984/28/24/244024}
       {J. Phys.: Condens. Matter {\bf 28}, 244024 (2016).}

\bibitem{schindler2019particleConserving}
  T. Schindler, R. Wittmann, and J. M. Brader,
  {Particle-conserving dynamics on the single-particle level,}
  \href{https://doi.org/10.1103/PhysRevE.99.012605}
       {Phys. Rev. E {\bf 99}, 012605 (2019).}


\bibitem{schmidt2018md}
  M. Schmidt,
  {Power functional theory for Newtonian many-body dynamics,}
  \href{https://doi.org/10.1063/1.5008608}
       {J. Chem. Phys. {\bf 148}, 044502 (2018).}

\bibitem{renner2022prl}
  J. Renner, M. Schmidt, and D. de las Heras, 
  {Shear and bulk acceleration viscosities in simple fluids,}
  \href{https://doi.org/10.1103/PhysRevLett.128.094502}
       {Phys. Rev. Lett. {\bf 128}, 094502 (2022).}

\bibitem{schmidt2015qpft}
  M. Schmidt,
  {Quantum power functional theory for many-body dynamics,}
  \href{https://doi.org/10.1063/1.4934881}
       {J. Chem. Phys. {\bf 143}, 174108 (2015).}

\bibitem{bruetting2019viscosity}
  M. Br\"utting, T. Trepl, D. de las Heras, and M. Schmidt,
  {Superadiabatic forces via the acceleration gradient in quantum 
    many-body dynamics,}
  \href{https://doi.org/10.3390/molecules24203660}
       {Molecules {\bf 24}, 3660 (2019).}


\bibitem{clegg2021ml}
  P. S. Clegg,
  {Characterising soft matter using machine learning},
  \href{https://doi.org/10.1039/D0SM01686A}
       {Soft Matter, {\bf 17}, 3991 (2021).}

\bibitem{dijkstra2021ml}
  M. Dijkstra and E. Luijten,
  {From predictive modelling to machine learning and reverse
    engineering of colloidal self-assembly},
  \href{https://doi.org/10.1038/s41563-021-01014-2}
       {Nature Materials {\bf 20}, 762 (2021).}

\bibitem{coli2022scienceAdvances}
  G. M. Coli, E. Boattini, L. Filion, and M. Dijkstra,
  {Inverse design of soft materials via a deep
    learning-based evolutionary strategy,}
  \href{https://doi.org/10.1126/sciadv.abj6731}
       {Sci. Adv. {\bf 8}, eabj6731 (2022).}

\bibitem{boattinia2019ml}
  E. Boattini, M. Dijkstra, and L. Filion,
  {Unsupervised learning for local structure detection in colloidal systems},
  \href{https://doi.org/10.1063/1.5118867}
       {J. Chem. Phys. {\bf 151}, 154901 (2019).}


\bibitem{mastrigt2022prl}
  R. van Mastrigt, M. Dijkstra, M. van Hecke, and C. Coulais,
  {Machine learning of implicit combinatorial rules in mechanical 
    metamaterials,}
  \href{https://doi.org/10.1103/PhysRevLett.129.198003}
       {Phys. Rev. Lett. {\bf 129}, 198003 (2022).}

\bibitem{campos2021ml}
  G. Campos-Villalobos, E. Boattini, L. Filion, and M. Dijkstra,
  {Machine learning many-body potentials for colloidal systems,}
  \href{https://doi.org/10.1063/5.0063377}
       {J. Chem. Phys. {\bf 155}, 174902 (2021).}

\bibitem{campos2022ml}
  G. Campos-Villalobos, G. Giunta, S. Marín-Aguilar, and M. Dijkstra,
  {Machine-learning effective many-body potentials for anisotropic 
    particles using orientation-dependent symmetry functions,}
  \href{https://doi.org/10.1063/5.0091319}
       {J. Chem. Phys. {\bf 157}, 024902 (2022).}

\bibitem{ciarella2022}
  S. Ciarella, M. Chiappini, E. Boattini, M. Dijkstra, and L. M. C. Janssen,
  {Dynamics of supercooled liquids from static averaged quantities using
    machine learning,}
  \href{https://doi.org/10.48550/arXiv.2212.09338}
       {arXiv:2212.09338.}

{\ms
\bibitem{winter2023ml}
  M. K. Winter, I. Pihlajamaa, V. E. Debets, and L. M. C. Janssen,
  A deep learning approach to the measurement of long-lived memory 
  kernels from generalised Langevin dynamics,
  \href{https://doi.org/10.48550/arXiv.2302.13682}
       {arXiv:2302.13682}.

\bibitem{janzen2023}
  G. Janzen, C. Smit, S. Visbeek, V. E. Debets, C. Luo, C. Storm, 
  S. Ciarella, and L. M.C. Janssen,
  Classifying the age of a glass based on structural properties: 
  A machine learning approach,
  \href{https://doi.org/10.48550/arXiv.2303.00636}
       {arXiv:2303.00636.}

\bibitem{paret2020}
  J. Paret, R. L. Jack, and D. Coslovich,
  Assessing the structural heterogeneity of supercooled
  liquids through community inference,
  \href{https://doi.org/10.1063/5.0004732}
       {J. Chem. Phys. {\bf 152}, 144502 (2020).}

\bibitem{coslovich2022}
  D. Coslovich, R. L. Jack, and J. Paret,
  Dimensionality reduction of local structure in glassy binary mixtures,
  \href{https://doi.org/10.1063/5.0128265}
       {J. Chem. Phys. {\bf 157}, 204503 (2022).}
       



\bibitem{singh2023ml}
  A. N. Singh and D. T. Limmer,
  Variational deep learning of equilibrium transition path ensembles,
  \href{https://doi.org/10.48550/arXiv.2302.14857}
       {arXiv:2302.14857}.

\bibitem{das2021}
  A. Das, D. C. Rose, J. P. Garrahan, and D. T. Limmer,
  Reinforcement learning of rare diffusive dynamics,
  \href{https://doi.org/10.1063/5.0057323}
       {J. Chem. Phys. {\bf 155}, 134105 (2021).}

\bibitem{lindquist2016}
  B. A. Lindquist, R. B. Jadrich, and T. M. Truskett,
  Communication: Inverse design for self-assembly via on-the-fly optimization,
  \href{https://doi.org/10.1063/1.4962754}
       {J. Chem. Phys. {\bf 145}, 111101 (2016).}

\bibitem{shermann2020}
  Z. M. Sherman, M. P. Howard, B. A. Lindquist, R. B. Jadrich, 
  and T. M. Truskett,
  Inverse methods for design of soft materials,
  \href{https://doi.org/10.1063/1.5145177}
       {J. Chem. Phys. {\bf 152}, 140902 (2020).}
       
\bibitem{statt2021}
  A. Statt, D. C. Kleeblatt, and  W. F. Reinhart,
  Unsupervised learning of sequence-specific aggregation behavior 
  for a model copolymer,
  \href{https://doi.org/10.1039/D1SM01012C}
       {Soft Matter {\bf 17}, 7697 (2021).}
     
\bibitem{mahynski2020}
  N. A. Mahynski, R. Mao, E. Pretti, V. K. Shena, and J. Mittal,
  Grand canonical inverse design of multicomponent colloidal crystals,
  \href{https://doi.org/10.1039/C9SM02426C}
       {Soft Matter {\bf 16}, 3187 (2020).}

\bibitem{oleary2021}
  J. O'Leary, R. Mao,  E. J. Pretti, J. A. Paulson, J. Mittal,
  Deep learning for characterizing the self-assembly of
  three-dimensional colloidal systems,
  \href{https://doi.org/10.1039/D0SM01853H }
       {Soft Matter {\bf 17}, 989 (2021).}

\bibitem{zhang2020}
  J. Zhang, J. Yang, Y. Zhang, and M. A. Bevan,
  Controlling colloidal crystals via morphing energy landscapes
  and reinforcement learning,
  \href{https://doi.org/10.1126/sciadv.abd6716}
       {Sci. Adv. {\bf 6}, eabd6716 (2020).}

\bibitem{sidky2018}
  H. Sidky and J. K. Whitmer,
  Learning free energy landscapes using artificial neural networks, 
  \href{https://doi.org/10.1063/1.5018708}
       {J. Chem. Phys. {\bf 148}, 104111 (2018).}

\bibitem{niblett2021}
  S. P. Niblett, M. Galib, and D. T. Limmer,
  Learning intermolecular forces at liquid-vapor interfaces,
  \href{https://doi.org/10.1063/5.0067565}
       {J. Chem. Phys. {\bf 155}, 164101 (2021).}

\bibitem{weeks1995}
  J. D. Weeks, R. L. B. Selinger, and J. Q. Broughton, 
  Self-consistent treatment of repulsive and attractive forces
  in nonuniform liquids,
  \href{https://doi.org/10.1103/physrevlett.75.2694}
       {Phys. Rev. Lett. {\bf 75}, 2694 (1995).}

\bibitem{weeks2002}
  J. D. Weeks,
  Connecting local structure to interface formation: A molecular
  scale van der Waals theory of nonuniform liquids,
  \href{https://doi.org/10.1146/annurev.physchem.53.100201.133929}
       {Annu. Rev. Phys. Chem. {\bf 53}, 533 (2002).}

\bibitem{archer2013lmft}
  A. J. Archer, and R. Evans,
  Relationship between local molecular field theory
  and density functional theory for nonuniform liquids,
  \href{https://doi.org/10.1063/1.4771976}
       {J. Chem. Phys. {\bf 138}, 014502 (2013).}
       

}


\bibitem{teixera2014}
  T. Santos-Silva, P. I. C. Teixeira, C. Anquetil-Deck, and D. J. Cleaver,
  {Neural-network approach to modeling liquid crystals in complex confinement,}
  \href{https://doi.org/10.1103/PhysRevE.89.053316}
       {Phys. Rev. E {\bf 89}, 053316 (2014).}

\bibitem{lin2019ml}
  S.-C. Lin and M. Oettel, 
  {A classical density functional from machine learning and a 
    convolutional neural network,}
  \href{http://dx.doi.org/10.21468/SciPostPhys.6.2.025}
       {SciPost Phys. {\bf 6}, 025 (2019).}

\bibitem{lin2020ml}
  S.-C. Lin,  G. Martius, and M. Oettel,
  {Analytical classical density functionals from an equation learning network,}
  \href{https://doi.org/10.1063/1.5135919}
       {J. Chem. Phys. {\bf 152}, 021102 (2020).}

\bibitem{cats2022ml}
  P. Cats, S. Kuipers, S. de Wind, R. van Damme, G. M. Coli,
  M. Dijkstra, and R. van Roij,
  {Machine-learning free-energy functionals using density 
    profiles from simulations,}
  \href{https://doi.org/10.1063/5.0042558}
       {APL Mater. {\bf 9}, 031109 (2021).}

{\ms
\bibitem{yatsyshin2022}
  P. Yatsyshin, S. Kalliadasis and A. B. Duncan,
  {Physics-constrained Bayesian inference of state functions 
  in classical density-functional theory,}
  \href{https://doi.org/10.1063/5.0071629}
       {J. Chem. Phys. {\bf 156}, 074105 (2022).}

\bibitem{fang2022}
  X. Fang, M. Gu and J. Wu,
  {Reliable emulation of complex functionals by active learning with 
  error control,}
  \href{https://doi.org/10.1063/5.0121805}
       {J. Chem. Phys. {\bf 157}, 214109 (2022).}

\bibitem{qiao2020}
  C. Qiao, X. Yu, X. Song, T. Zhao, X. Xu, S. Zhao, and K. E. Gubbins,
  Enhancing gas solubility in nanopores: a combined study using
  classical density functional theory and machine learning,
  \href{https://doi.org/10.1021/acs.langmuir.0c01160}
       {Langmuir {\bf 36}, 8527 (2020).}


}



\bibitem{rotenberg2020}
  B. Rotenberg,
  {Use the force! Reduced variance estimators for densities,
    radial distribution functions, and local mobilities in
    molecular simulations,}
  \href{https://doi.org/10.1063/5.0029113}
       {J. Chem. Phys. {\bf 153}, 150902 (2020).}

\bibitem{borgis2013}
  D. Borgis, D., R. Assaraf, B. Rotenberg, and R. Vuilleumier,
  {Computation of pair distribution functions and three-dimensional
    densities with a reduced variance principle,}
  \href{https://doi.org/10.1080/00268976.2013.838316}
       {Mol. Phys. {\bf 111}, 3486 (2013).}

\bibitem{delasheras2018forceSampling}
  D. de las Heras and M. Schmidt,
  {Better than counting: Density profiles from force sampling,}
  \href{https://doi.org/10.1103/PhysRevLett.120.218001}
       {Phys. Rev. Lett. {\bf 120}, 218001 (2018).}

\bibitem{renner2023torqueSampling}
  J. Renner, M. Schmidt, and D. de las Heras,
  {Better than counting: Orientational distribution functions 
    from torque sampling,}
  \href{https://doi.org/10.48550/arXiv.2212.11576}
       {arXiv:2212.11576}.

\bibitem{he2021}
  B. He, I. Martin-Fabiani, R. Roth, G. I. T\'oth, and A. J. Archer,
  {Dynamical density functional theory for the drying and
    stratification of binary colloidal dispersions,}
  \href{https://doi.org/10.1021/acs.langmuir.0c02825}
       {Langmuir {\bf 37}, 1399 (2021). }

\bibitem{kundu2022ddft}
  M. Kundu and M. P. Howard,
  {Dynamic density functional theory for drying colloidal
  suspensions: Comparison of hard-sphere free-energy functionals,}
  \href{https://doi.org/10.1063/5.0118695}
     {J. Chem. Phys. {\bf 157}, 184904 (2022).}

\bibitem{sui2018}
  J. Sui, M. Doiab  and  Y. Ding,
  {Dynamics of the floating nematic phase formation in platelet
    suspension with thickness polydispersity by sedimentation,}
  \href{https://doi.org/10.1039/C8SM01177J}
       {Soft Matter {\bf 14}, 8956 (2018).}


\bibitem{tschopp2022ddft2}
  S. M. Tschopp and J.~M. Brader,
  {First-principles superadiabatic theory for the dynamics of inhomogeneous 
    fluids},
  \href{https://doi.org/10.1063/5.0131441}
       {J. Chem. Phys. {\bf 157}, 234108 (2022).}



{\ms
\bibitem{scacchi2016}
  A. Scacchi, M. Kr\"uger and J. M. Brader,
  Driven colloidal fluids: construction of dynamical density
  functional theories from exactly solvable limits,
  \href{doi:10.1088/0953-8984/28/24/244023}
       {J. Phys.: Condens. Matter {\bf 28}, 244023 (2016).}
}

\bibitem{molinero2009}
  V. Molinero and E. B. Moore,
  {Water modeled as an intermediate element between
    carbon and silicon,}
  \href{https://doi.org/10.1021/jp805227c}
       {J. Phys. Chem. B {\bf 113}, 4008–4016 (2009).}

\bibitem{coe2022water}
  M. K. Coe, R. Evans, and N. B. Wilding,
  {The coexistence curve and surface tension of
    a monatomic water model,}
  \href{https://doi.org/10.1063/5.0085252}
       {J. Chem. Phys. {\bf 156}, 154505 (2022).}

\bibitem{saw2009}
  S. Saw, N. L. Ellegaard, W. Kob, and S. Sastry,
  {Structural relaxation of a gel modeled by three body interactions,}
  \href{https://doi.org/10.1103/PhysRevLett.103.248305}
       {Phys. Rev. Lett. {\bf 103}, 248305 (2009).}

\bibitem{saw2011}
  S. Saw, N. L. Ellegaard, W. Kob, and S. Sastry,
  {Computer simulation study of the phase behavior and structural
    relaxation in a gel-former modeled by three-body interactions,}
  \href{https://doi.org/10.1063/1.3578176}
       {J. Chem. Phys. {\bf 134}, 164506 (2011).}



\bibitem{doi2011}
  M. Doi,
  {Onsager's variational principle in soft matter,}
  \href{https://doi.org/10.1088/0953-8984/23/28/284118}
       {J. Phys.: Condens. Matter {\bf 23}, 284118 (2011).}

\bibitem{doi2015}
  M. Doi,
  {Onsager principle as a tool for approximation,}
  \href{https://doi.org/10.1088/1674-1056/24/2/020505}
       {Chinese Phys. B {\bf 24}, 020505 (2015).}

\bibitem{wang2021onsager}
  H. Wang, T. Qian, and X. Xu,
  {Onsager's variational principle in active soft matter,}
  \href{https://doi.org/10.1039/d0sm02076a}
       {Soft Matter {\bf 17}, 3634 (2021).}

\bibitem{wang2022prl}
  X. Wang, J. Dobnikar, and D. Frenkel,
  {Numerical test of the Onsager relations in a driven system},
  \href{https://doi.org/10.1103/PhysRevLett.129.238002}
       {Phys. Rev. Lett. {\bf 129}, 238002 (2022).}

\bibitem{seifert2012}
  U. Seifert,
  {Stochastic thermodynamics, fluctuation theorems and molecular machines,}
  \href{https://doi.org/10.1088/0034-4885/75/12/126001}
       {Rep. Prog. Phys. {\bf 75}, 126001 (2012).}

\bibitem{jack2010}
  R. L. Jack and P. Sollich,
  {Large Deviations and Ensembles of Trajectories in Stochastic Models,}
  \href{https://doi.org/10.1143/PTPS.184.304}
       {Prog. Theo. Phys. Suppl. {\bf 184}, 304 (2010).}

\bibitem{jack2015}
  R. L. Jack and P. Sollich,
  {Effective interactions and large deviations in stochastic processes,}
  \href{https://doi.org/10.1140/epjst/e2015-02416-9}
       {Europ. Phys. J. Spec. Top. {\bf 224}, 2351 (2015).}

\bibitem{janssen2018}
  L. M. C. Janssen,
  {Mode-coupling theory of the glass transition: a primer,}
  \href{https://doi.org/10.3389/fphy.2018.00097}
       {Front. Phys. {\bf 6}, 97 (2018).}

\bibitem{janssen2015}
  L. M. C. Janssen and D. R. Reichman,
  {Microscopic dynamics of supercooled liquids from first principles,}
  \href{http://dx.doi.org/10.1103/PhysRevLett.115.205701}
       {Phys. Rev. Lett. {\bf 115}, 205701 (2015).}

\bibitem{mazzuca2023}
  G. Mazzuca, T. Grava, T. Kriecherbauer, K. T.-R. McLaughlin, 
  C. B. Mendl, and H. Spohn,
  {Equilibrium spacetime correlations of the toda lattice on the 
    hydrodynamic scale,}
  \href{https://doi.org/10.48550/arXiv.2301.02431}
       {arXiv:2301.02431.}

{\ms
\bibitem{rodgers2020}
  E. B. Baker, J. M. Rodgers, and J. D. Weeks,
  Local molecular field theory for nonequilibrium systems,
  \href{https://dx.doi.org/10.1021/acs.jpcb.0c03295}
       {J. Phys. Chem. B {\bf 124}, 5676 (2020).}

}

\bibitem{lips2018}
  D. Lips, A. Ryabov, and P. Maass,
  {Brownian asymmetric simple exclusion process,}
  \href{https://doi.org/10.1103/PhysRevLett.121.160601}
       {Phys. Rev. Lett. {\bf 121}, 160601 (2018).}

\bibitem{antonov2022}
  A. P. Antonov, A. Ryabov, and P. Maass,
  {Solitons in overdamped Brownian dynamics,}
  \href{https://doi.org/10.1103/PhysRevLett.129.080601}
       {Phys. Rev. Lett. {\bf 129}, 080601 (2022).}

{\ms

\bibitem{haussmann2016}
  R. Haussmann,
  The way from microscopic many-particle theory to macroscopic hydrodynamics,
  \href{https://doi.org/10.1088/0953-8984/28/11/113001}
       {J. Phys.: Condens. Matter {\bf 28}, 113001 (2016).}

\bibitem{haussmann2022}
  R. Haussmann,
  Microscopic density-functional approach to nonlinear elasticity theory,
  \href{https://doi.org/10.1088/1742-5468/ac6d61}
       {J. Stat. Mech. {\bf 2022}, 053210 (2022).}

\bibitem{ganguly2022}
  S. Ganguly, G. P. Shrivastav, S.-C. Lin, J. H\"aring,
  R. Haussmann, G. Kahl, M. Oettel, and M. Fuchs
  Elasticity in crystals with a high density of local defects:
  Insights from ultra-soft colloids
  \href{https://doi.org/10.1063/5.0073624}
       {J. Chem. Phys. {\bf 156}, 064501 (2022).}

\bibitem{lin2021dcf}
  S.-C. Lin, M. Oettel, J. H\"aring, R. Haussmann, M. Fuchs, and G. Kahl,
  The direct correlation function of a crystalline solid,
  \href{https://doi.org/10.1103/physrevlett.127.085501}
      {Phys. Rev. Lett. {\bf 127}, 085501 (2021).}

\bibitem{anero2013}
  J. G. Anero, P. Espa\~nol, and P. Tarazona,
  Functional thermo-dynamics: A generalization of dynamic density
  functional theory to non-isothermal situations, 
  \href{ https://doi.org/10.1063/1.4811655}
       {J. Chem. Phys. {\bf 139}, 034106 (2013).}

\bibitem{zwanzig2001}
  R. Zwanzig, 
  {\it Nonequilibrium Statistical Mechanics}
  (Oxford University Press, Oxford, 2001).

\bibitem{vanhove1954}
 L. van Hove, 
 Correlations in space and time and Born approximation scattering in
 systems of interacting particles, 
 \href{https://doi.org/10.1103/PhysRev.95.249}
      {Phys. Rev. {\bf 95}, 249 (1954).}

\bibitem{brader2013noz}
  J. M. Brader and M. Schmidt, 
  Dynamic correlations in Brownian many-body systems,
  \href{http://dx.doi.org/10.1063/1.4861041}
       {J. Chem. Phys. {\bf 140}, 034104 (2014).}

\bibitem{brader2014noz}
  J. M. Brader and M. Schmidt, 
  Nonequilibrium Ornstein-Zernike relation for Brownian many-body dynamics,
  \href{http://dx.doi.org/10.1063/1.4820399}
       {J. Chem. Phys. {\bf 139}, 104108 (2013).}

\bibitem{mermin1965}
  N. D. Mermin,
  Thermal properties of the inhomogeneous electron gas,
  \href{https://doi.org/10.1103/PhysRev.137.A1441}
       {Phys. Rev. {\bf 137}, A1441 (1965).}


\bibitem{vanleeuwen1999}
  R. van Leeuwen,
  Mapping from densities to potentials in time-dependent density-functional theory,
  \href{https://doi.org/10.1103/PhysRevLett.82.3863}
       {Phys. Rev. Lett. {\bf 82}, 3863 (1999).}

\bibitem{maitra2002}
  N. T. Maitra, K. Burke, and C. Woodward,
  Memory in time-dependent density functional theory
  \href{https://doi.org/10.1103/PhysRevLett.89.023002}
       {Phys. Rev. Lett. {\bf 89}, 023002 (2002).}

\bibitem{zhao2020}
  H. Zhao, B. D. Storey, R. D. Braatz, and M. Z. Bazant,
  Learning the physics of pattern formation from images,
  \href{https://doi.org/10.1103/PhysRevLett.124.060201}
       {Phys. Rev. Lett. {\bf 124}, 060201 (2020).}

\bibitem{bhatio2021}
  H. Bhatia, T. S. Carpenter, H. I. Ing\'olfsson, G. Dharuman, P. Karande, 
  S. Liu, T. Oppelstrup, C. Neale, F. C. Lightstone, B. Van Essen, 
  J. N. Glosli, and P.-T. Bremer,
  Machine-learning-based dynamic-importance sampling
  for adaptive multiscale simulations,
  \href{https://doi.org/10.1038/s42256-021-00327-w}
       {Nat. Mach. Intel. {\bf 3}, 401 (2021).}


\bibitem{schmid2022editorial}
  F. Schmid,
  Editorial: Multiscale simulation methods for soft matter systems,
  \href{https://doi.org/10.1088/1361-648X/ac5071}
       {J. Phys.: Condens. Matter {\bf 34}, 160401 (2022).}

\bibitem{ebrahimi2021}
  R. Ebrahimi Viand, F. H\"ofling, R. Klein, and L. Delle Site,
  Theory and simulation of open systems out of equilibrium,
  \href{https://doi.org/10.1063/5.0014065}
       {J. Chem. Phys. {\bf 153}, 101102 (2021).}

\bibitem{dellesitte2019}
  L. Delle Site, C. Krekeler, J. Whittaker, A. Agarwal, R. Klein, 
  and F. H\"ofling,
  Molecular Dynamics of open systems: construction of a mean‐field 
  particle reservoir,
  \href{ https://doi.org/10.1002/adts.201900014}
       {Adv. Theo.  Sim. {\bf 2}, 1900014 (2019)}.

\bibitem{baptista2021}
 L. A. Baptista, R. C. Dutta, M. Sevilla, M. Heidari, R. Potestio, 
 K. Kremer, and R. Cortes-Huerto,
 Density-functional-theory approach to the
 Hamiltonian adaptive resolution simulation method,
 \href{https://doi.org/10.1088/1361-648X/abed1d}
     {J. Phys.: Condens. Matter {\bf 33}, 184003 (2021).}

\bibitem{gholami2021}
  A. Gholami, F H\"ofling, R. Klein, and L. Delle Site,
  Thermodynamic relations at the coupling boundary in
  adaptive resolution simulations for open systems,
  \href{https://doi.org/10.1002/adts.202000303}
       {Adv. Theo. Sim. {\bf 4}, 2000303 (2021).}

\bibitem{klippenstein2021}
  V. Klippenstein, M. Tripathy, G. Jung, F. Schmid, and N. F. A. van der Vegt,
  Introducing memory in coarse-grained molecular simulations,
  \href{https://doi.org/10.1021/acs.jpcb.1c01120}
       {J. Phys. Chem. B {\bf 125}, 4931 (2021).}

\bibitem{jung2017reconstructing}
  G. Jung, M. Hanke, and F. Schmid,
  Iterative reconstruction of memory kernels
  \href{https://doi.org/10.1021/acs.jctc.7b00274}
       {J. Chem. Theo. Comp. {\bf 13}, 2481 (2017).}

\bibitem{tchenkoue2019}
  M.-L. M. Tchenkoue, M. Penz, I. Theophilou, M. Ruggenthaler, and A. Rubio,
  Force balance approach for advanced approximations in density functional theories,
  \href{https://doi.org/10.1063/1.5123608}
       {J. Chem. Phys. {\bf 151}, 154107 (2019).}



\bibitem{tarantino2021}
  W. Tarantino and C. A. Ullrich,
  A reformulation of time-dependent Kohn-Sham theory in terms
  of the second time derivative of the density,
  \href{https://doi.org/10.1063/5.0039962}
      {J. Chem. Phys. {\bf 154}, 204112 (2021).}


\bibitem{tokatly2005one}
  I. V. Tokatly,
  Quantum many-body dynamics in a Lagrangian frame: I. Equations of motion
  and conservation laws,
  \href{https://doi.org/10.1103/PhysRevB.71.165104}
       {Phys. Rev. B {\bf 71}, 165104 (2005).}

\bibitem{tokatly2005two}
  I. V. Tokatly,
  Quantum many-body dynamics in a Lagrangian frame: II. Geometric formulation
  of time-dependent density functional theory,
  \href{https://doi.org/10.1103/PhysRevB.71.165105}
       {Phys. Rev. B {\bf 71}, 165105 (2005).}

\bibitem{ullrich2006}
  C. A. Ullrich and I. V. Tokatly,
  Nonadiabatic electron dynamics in time-dependent density-functional theory,
  \href{https://doi.org/10.1103/PhysRevB.73.235102}
       {Phys. Rev. B {\bf 73}, 235102 (2006).}


\bibitem{tokatly2007}
  I. V. Tokatly,
  Time-dependent deformation functional theory,
  \href{https://doi.org/10.1103/PhysRevB.75.125105}
       {Phys. Rev. B {\bf 75}, 125105 (2007).}


}

\end{thebibliography}
\end{document}